\documentclass[aps,prc,superscriptaddress,twocolumn,showpacs,nofootinbib]{revtex4-1}
\usepackage[utf8]{inputenc}
\usepackage[T1]{fontenc}
\usepackage{graphicx}   
\usepackage{latexsym}   
\usepackage{enumerate}
\usepackage{rotating,booktabs,multirow}
\usepackage{amsmath}
\usepackage{amsfonts}
\usepackage{amssymb}
\usepackage{multirow}
\usepackage[normalem]{ulem}
\usepackage{bm}
\usepackage{subfigure}
\usepackage{tikz}
\usepackage{xcolor}
\usepackage[left]{lineno} 
\usepackage[colorlinks=true,linktocpage=true,linkcolor=blue,citecolor=blue,allcolors=blue]{hyperref}

\definecolor{lime}{HTML}{A6CE39}
\DeclareRobustCommand{\orcidicon}{
	\begin{tikzpicture}
	\draw[lime, fill=lime] (0,0) 
	circle [radius=0.16] 
	node[white] {{\fontfamily{qag}\selectfont \tiny ID}};
	\draw[white, fill=white] (-0.0625,0.095) 
	circle [radius=0.007];
	\end{tikzpicture}
	\hspace{-2mm}
}
\foreach \x in {A, ..., Z}{\expandafter\xdef\csname orcid\x\endcsname{\noexpand\href{https://orcid.org/\csname orcidauthor\x\endcsname}
			{\noexpand\orcidicon}}
}

\begin{document}

\title{Robustness of the scaled transverse-momentum correlator and the
       contribution of meson--baryon fluctuations at RHIC energies}

\author{Rutik Manikandhan\orcidA{}}
\affiliation{Department of Physics, The Ohio State University, Columbus, OH 43210, USA}

\author{Rene Bellwied\orcidC{}}
\affiliation{Department of Physics, University of Houston, Houston, TX 77204, USA}

\begin{abstract}
The search for the QCD critical point relies on identifying non-monotonic signatures in event-by-event fluctuation observables, such as the scaled two-particle transverse-momentum correlator $C_{p_T}$. Using a microscopic transport model (UrQMD), we establish the robustness of $C_{p_T}$ against the variations that distinguish the datasets in the relevant collision energy regime, namely the choice of centrality estimator, finite tracking efficiency, and the acceptance, including the difference between collider and fixed-target running. In addition, it has recently been proposed, using UrQMD yields, that the measured non-monotonicity requires no critical dynamics and follows instead from the event-by-event composition of the final state. We show that this contribution is governed by event-by-event fluctuations in the meson–baryon composition, rather than by the mean yields alone. Using the composition fluctuations obtained from UrQMD, we find that the minimum in the two-component calculation disappears.

\end{abstract}

\maketitle

\section{Introduction}
The exploration of the phase structure of strongly interacting matter and the search for the onset of deconfinement and the QCD critical point remain central goals of relativistic heavy-ion collision experiments~\cite{Stephanov:2004wx,Asakawa:2000wh}. First-principles lattice QCD calculations indicate that at vanishing baryon chemical potential ($\mu_B = 0$) the transition from hadronic matter to a deconfined quark--gluon plasma is a smooth crossover~\cite{Borsanyi:2010bp,Bazavov:2011nk}. However, extending these calculations to finite $\mu_B$ is hindered by the well-known sign problem, leading to significant uncertainties in the determination of the phase boundary~\cite{deForcrand:2009zkb,Muroya:2003qs}. Various theoretical approaches suggest that at sufficiently large baryon densities the transition becomes first order, implying the existence of a critical endpoint separating the crossover and first-order regions of the QCD phase diagram~\cite{Shah_2026, cbwj-4jfl, Stephanov:1998dy,Fodor:2004nz}, although alternative scenarios with a smooth crossover at all $\mu_B$ have also been discussed~\cite{Aarts:2015tyj}.

Experimentally, heavy-ion collisions over a broad range of beam energies provide access to different regions of the QCD phase diagram. Ongoing and completed experimental programmes, including the Beam Energy Scan at RHIC and the NA61/SHINE~\cite{NA61SHINE:2023unn,Stefanek:2026ltn} programme at the CERN SPS, have investigated the properties of strongly interacting matter and searched for signatures of the phase transition and the QCD critical point. Future facilities, such as the Compressed Baryonic Matter (CBM) experiment at FAIR, will extend these studies into the high-baryon-density region of the phase diagram~\cite{CBM:2016kpk}. In this context, event-by-event observables, including fluctuations and correlations, are sensitive to the underlying dynamics and to the correlation length of the created medium~\cite{Stephanov:1999zu,Koch:2025cog}.

While higher-order cumulants of conserved charges have been extensively studied as probes of critical phenomena~\cite{Asakawa:2000wh,Jeon:2000wg,STAR:2025ydk,STAR:2025zdq}, transverse-momentum ($p_T$) correlations provide a complementary probe of the medium. Event-by-event $p_T$ correlations are sensitive to both dynamical fluctuations and collective behaviour, offering insight into temperature fluctuations, initial-state density inhomogeneities, and the transport properties of the medium~\cite{Voloshin:1999yf,Gavin:2003cb,Gavin:2006xd,liu2025newconstraintsequationstate}. Beyond their potential sensitivity to critical phenomena, observables based on the event-by-event mean transverse momentum have also been used to constrain the thermodynamic properties of the produced matter. The correlation between $\langle p_{\rm T}\rangle$ and charged-particle multiplicity in ultra-central collisions has been used to extract the speed of sound~\cite{2024,ALICE:2025pTmoments}, while the radial-flow coefficient $v_{0}(p_{\rm T})$, which characterises the response of the transverse-momentum spectrum to event-by-event fluctuations of the mean, constrains the shear and bulk viscosities~\cite{ldcn-r2lq,9pvq-ph1c}. These applications highlight the potential of $p_T$-based observables to probe the properties of the medium. Recently, the STAR Collaboration reported a non-monotonic beam-energy dependence of transverse-momentum correlations in Au+Au collisions, with a significance of approximately $5\sigma$~\cite{STAR:2026vjv}. This measurement motivates a systematic investigation of how experimental effects and analysis choices influence the measured observable.

The STAR measurement combines data collected over multiple years and under different experimental configurations, including detector upgrades, collider and fixed-target running, variations in tracking performance, and changes in centrality determination. These differences can affect the measured correlations through variations in detector efficiency and event selection, as well as through the phase-space regions used to define centrality~\cite{Luo:2013bmi,Luo:2011ts,Bzdak:2012an}.

Reichert \textit{et al.}~\cite{Reichert:2026mbm} proposed that variations in the relative contributions of mesons and baryons, meson--baryon mixing, could account for the observed non-monotonicity.

In this work, we quantify the sensitivity of $p_T$ correlations to centrality definition, detector efficiency, and acceptance, and examine the proposed meson--baryon mixing. These studies establish the robustness of the measured non-monotonicity against the experimental effects considered and assess whether meson--baryon mixing can explain the observed behaviour.

\section{Observable}
Transverse momentum correlations are a sensitive probe of the thermodynamic properties of the hot and dense QCD medium produced in heavy-ion collisions~\cite{Stodolsky:1995ds,Cao:2021zhy,Basu:2016ibk}. In particular, dynamical fluctuations of the event-by-event mean transverse momentum are expected to reflect temperature and energy fluctuations in the system and are related to its heat capacity near the conjectured QCD critical point~\cite{Stodolsky:1995ds,Basu:2016ibk}.

We use the two-particle transverse-momentum correlator in the form employed by the STAR Collaboration~\cite{STAR:2019clv,STAR:2026vjv},
\begin{equation}
\begin{split}
\langle \Delta p_{T,i}\,\Delta p_{T,j}\rangle
&=\frac{1}{N_{\mathrm{ev}}}\sum_{k=1}^{N_{\mathrm{ev}}}
  \frac{1}{N_k\left(N_k-1\right)}\\
&\quad\times\sum_{i=1}^{N_k}\sum_{\substack{j=1\\ j\neq i}}^{N_k}
  \delta p_{T,i}\,\delta p_{T,j},
\end{split}
\label{eq:corr}
\end{equation}
where $\delta p_{T,i}\equiv p_{T,i}-\langle\langle p_{\rm T}\rangle\rangle$,
$N_k$ is the number of accepted charged particles in event $k$, and
\begin{equation}
\langle\langle p_{\rm T}\rangle\rangle
=\frac{1}{N_{\mathrm{ev}}}\sum_{k=1}^{N_{\mathrm{ev}}}
\frac{1}{N_k}\sum_{i=1}^{N_k} p_{T,i}
\label{eq:ppt}
\end{equation}
is the inclusive mean transverse momentum of the centrality class, formed as an average over events.

The distinction between $\langle\langle p_{\rm T}\rangle\rangle$ and the event-wise mean $\langle p_{\rm T}\rangle_{k}$ is essential. Subtracting the event-wise mean would reduce Eq.~(\ref{eq:corr}) identically to $-\,\sigma^{2}_{p_T,k}/(N_k-1)$, a quantity that is negative by construction and carries no information about dynamical correlations. Subtracting the inclusive mean instead makes the correlator vanish exactly for statistically independent particle emission, so that a non-zero value measures genuine event-by-event correlations.

Expanding the product in Eq.~(\ref{eq:corr}) and averaging over events gives the equivalent form used in the present analysis,
\begin{equation}
\langle \Delta p_{T,i}\,\Delta p_{T,j}\rangle
=\bigl\langle C_{m1}\bigr\rangle-\langle\langle p_{\rm T}\rangle\rangle^{2},
\label{eq:cm1}
\end{equation}
with the event-wise pair average
\begin{equation}
C_{m1,k}=\frac{1}{N_k(N_k-1)}\sum_{i\neq j} p_{T,i}\,p_{T,j}.
\label{eq:cm1def}
\end{equation}
This form is numerically convenient because both terms are accumulated event by event; it is algebraically identical to Eq.~(\ref{eq:corr}).

The correlator is made dimensionless by scaling with the inclusive mean transverse momentum. The scaled correlator conventionally quoted in the literature is
\begin{equation}
C_{p_T}=\frac{\sqrt{\langle \Delta p_{T,i}\,\Delta p_{T,j}\rangle}}{\langle\langle p_{\rm T}\rangle\rangle},
\label{eq:CpT}
\end{equation}
and we use it throughout, so that the results can be compared directly with published measurements~\cite{ALICE:2014ccw,STAR:2019clv,Appelshauser:2004xj,PhysRevLett.133.252301}.

Statistical uncertainties are estimated using the bootstrap method, which accounts for statistical fluctuations in the measured quantities and their correlations. This approach provides a robust estimate of the statistical uncertainty without assuming that $\langle C_{m1}\rangle$ and $\langle\langle p_{\rm T}\rangle\rangle$ are independent.

\section{Transport codes}

Microscopic transport models provide the standard non-critical reference against which fluctuation and correlation observables in heavy-ion collisions are judged. In these approaches the collision is followed as an
explicit space-time evolution of individual degrees of freedom, with particle production, rescattering and decay governed by measured or effective cross sections, and with energy, momentum, baryon number, strangeness and electric charge conserved event by event. Because none of
them contains a critical point or a first-order phase boundary, any structure they generate in a correlation observable must arise from ordinary hadronic dynamics, resonance formation and decay, rescattering, conservation laws and the changing chemical composition of the final state rather than from critical phenomena.

Several such codes are available and are used interchangeably for this purpose in the field. Purely hadronic cascades such as UrQMD~\cite{Bass:1998ca,Bleicher_1999}, SMASH~\cite{Weil:2016zrk} and JAM~\cite{Nara:1999dz} share the same basic transport framework and differ mainly in their treatment of resonances, potentials and string
fragmentation. Models with an explicit partonic stage, such as AMPT~\cite{Lin:2004en} and PHSD~\cite{Cassing:2009vt}, add parton transport and hadronisation on top of an initial-state generator, while
HIJING~\cite{Wang:1991hta} used, for example, as the reference in
transverse-momentum fluctuation measurements at the
LHC~\cite{ALICE:2014ccw} and EPOS~\cite{Werner:2010aa} provide initial-state and full event descriptions over a wide range of collision energies. Several of these have already been confronted with the correlator considered here: AMPT calculations of the scaled transverse-momentum correlator have been published~\cite{TODO:ampt_prc}
and are used as one of the model comparisons accompanying the STAR measurement~\cite{STAR:2026vjv}, and a comparison based on the JAM transport model has appeared more recently~\cite{liu2026meanptfluctuationsauaucollisions}. This study complements previous comparisons of model predictions with the measured excitation function by examining the sensitivity of the observable to analysis choices. We test the robustness of the scaled transverse-momentum correlator against variations in centrality estimator, detector efficiency, and acceptance. These tests are not specific to a particular event generator and can be performed with any model that provides a realistic final state and respects relevant conservation laws.

For the present study we use the Ultra-relativistic Quantum Molecular Dynamics model\footnote{The model and its documentation are available at \url{urqmd.org}.} (UrQMD), version~3.4~\cite{Bass:1998ca,Bleicher_1999,%
Bleicher:2022kcu}. The code is open source and freely distributed, it is well documented and straightforward to compile and run, it covers the full beam-energy range of interest without retuning, and it has already been used as the non-critical baseline for the scaled transverse-momentum correlator in an earlier work~\cite{STAR:2019clv}, so that the results shown here connect directly to that body of results.

UrQMD is a microscopic hadronic transport model: hadrons propagate along classical trajectories and interact through binary collisions, resonance formation and resonance decay, with the interaction probability set by the geometric interpretation of the total cross section. Cross sections are taken from measured data where these exist, and are otherwise obtained from detailed balance or from effective-model input. The model treats the
established hadron species and their resonances up to masses of roughly $2.2$~GeV, together with their antiparticles and isospin partners, in line
with the PDG~\cite{ParticleDataGroup:2024cfk}. Above centre-of-mass energies of a few~GeV per nucleon pair, particle production proceeds
predominantly through the excitation and subsequent fragmentation of colour strings. As a QMD-type model~\cite{PhysRevC.37.2451,Aichelin:1991xy},
UrQMD represents nucleons by Gaussian wave packets whose centroids follow Hamilton's equations of motion,
$\dot{\mathbf{r}}_i = \partial \langle \mathcal{H} \rangle / \partial \mathbf{p}_i$
and
$\dot{\mathbf{p}}_i = -\,\partial \langle \mathcal{H} \rangle / \partial \mathbf{r}_i$,
so that potential interactions can be introduced directly into the
propagation. All $n$-body correlations generated by the collision dynamics
are retained.

The tests of Secs.~\ref{sec:cent}--\ref{sec:acceptance} are performed at
$\sqrt{s_{NN}} = 7.7$~GeV, where approximately $50$~million minimum-bias
Au+Au events were generated. This is the energy at which the measured
excitation function has its minimum, and at which its departure from the
monotonic trend of the higher energies is most
significant~\cite{STAR:2026vjv}; it is therefore the energy at which the
robustness of the observable matters most. It is also the lowest energy of
the collider programme, where the charged-particle multiplicity, and with it
the centrality resolution, is smallest. The tests concern the estimator and
the analysis procedure rather than the dynamics at a particular energy, and
their conclusions therefore apply equally at the other energies of the scan.

Unless stated otherwise, all events are generated in the centre-of-mass
(collider) frame, in cascade mode and without mean-field potentials, using
the default parameter set.

\section{Robustness of $C_{p_T}$ against the centrality estimator}
\label{sec:cent}

A reliable evaluation of dynamical transverse-momentum correlations
requires that the observable be free from biases introduced by the
centrality selection procedure. Centrality
in a heavy-ion experiment is not measured directly but inferred from the charged-particle multiplicity in some region of the detector. If that region overlaps the acceptance in which the correlator is evaluated, then
the same particles both define the centrality class and enter the observable, and an upward fluctuation in multiplicity simultaneously shifts the event into a more central class and alters the measured $\langle p_T \rangle$ of that event. In this section we test the robustness of $C_{p_T}$ against this effect within the UrQMD framework~\cite{Bleicher:2022kcu,Bass:1998ca}.

To this end we employ three centrality estimators. Two of them, RefMult and RefMult2, follow the reference-multiplicity conventions established in STAR analyses of charged-particle multiplicity fluctuations~\cite{STAR:2003oku} and used throughout the beam-energy scan~\cite{STAR:2020tga}: RefMult counts charged particles within
$|\eta| < 0.5$ in the centre-of-mass frame, and RefMult2 counts them in
the non-overlapping interval $0.5 < |\eta| < 1.0$. The correlator itself
is evaluated over $|\eta| < 0.5$. The two estimators therefore sit at
opposite extremes of the effect described above: RefMult is constructed
from exactly the particles that enter the correlator, so any
auto-correlation bias is present at full strength, whereas RefMult2 shares
no particle with the analysis acceptance and is free of it by
construction. A difference between the two would be a direct measure of
the bias. The third estimator is the impact parameter $b$, available in
the model but not in experiment, which provides a purely geometric
selection that is independent of final-state particle production
altogether and thus serves as the reference for the
multiplicity-based definitions.

\begin{figure}[htbp]
\centering
\includegraphics[width=\linewidth, trim=5 5 5 5, clip]{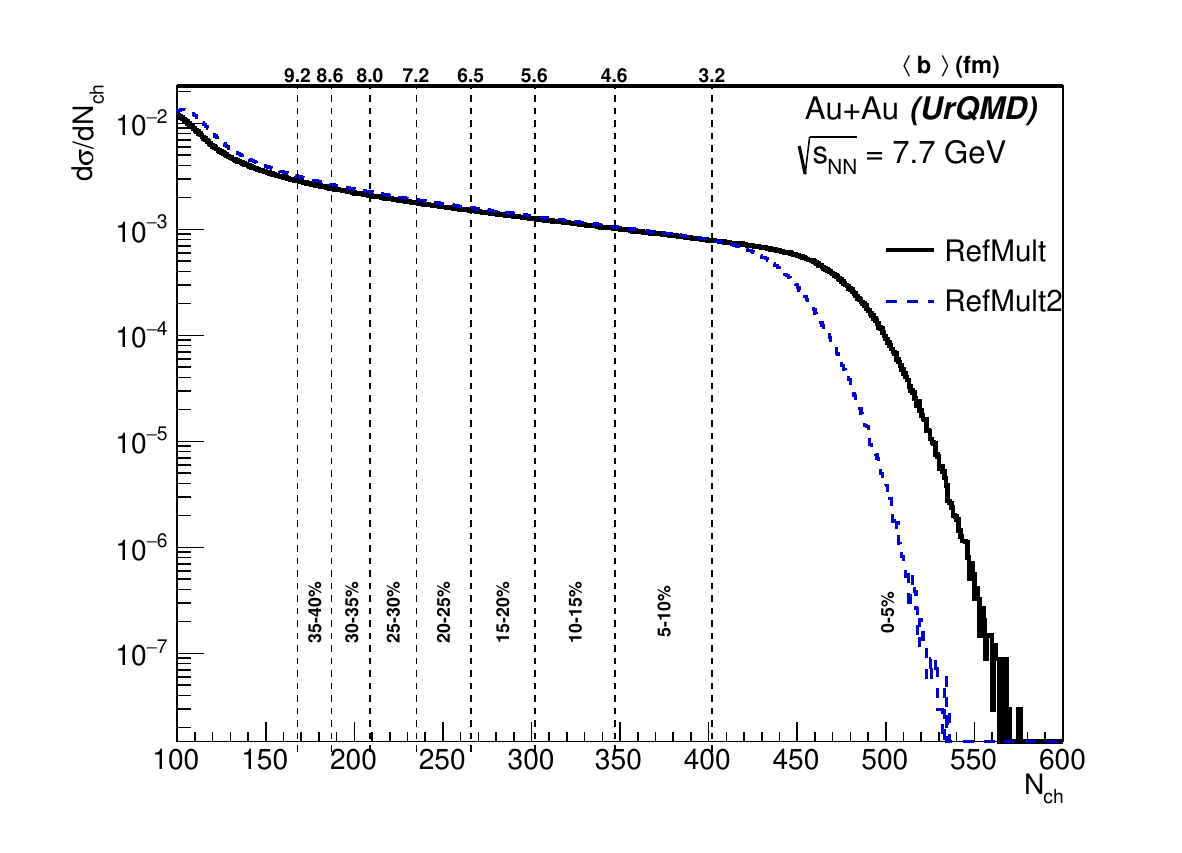}
\caption{Distributions of RefMult (black) and RefMult2 (blue) in Au+Au
collisions at $\sqrt{s_{\text{NN}}}=7.7~\mathrm{GeV}$. The vertical dashed
lines indicate the centrality class boundaries for RefMult.}
\label{fig:refmult_dist}
\end{figure}

The distributions of RefMult and RefMult2 for Au+Au collisions at
$\sqrt{s_{\mathrm{NN}}} = 7.7$~GeV are shown in
Fig.~\ref{fig:refmult_dist}. Both exhibit the expected behaviour for
heavy-ion collisions and are used to define centrality classes
consistently across the different estimators.

\begin{figure}[htbp]
\centering
\includegraphics[width=\linewidth]{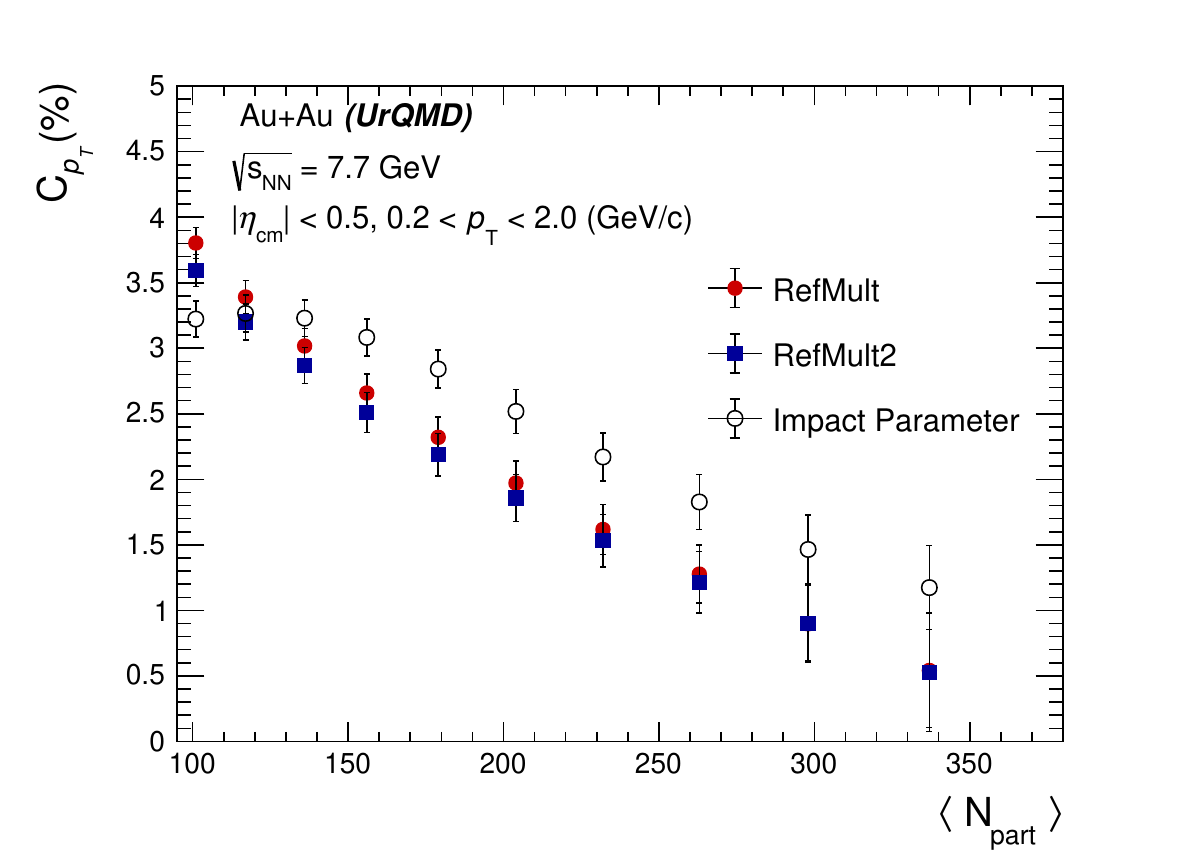}
\caption{Transverse-momentum correlator $C_{p_T}$ as a function of
centrality in Au+Au collisions at $\sqrt{s_{\text{NN}}}=7.7~\mathrm{GeV}$.
Results are shown for three centrality definitions: RefMult (red solid
circles), RefMult2 (blue solid circles), and the impact parameter $b$
(black hollow circles). Statistical uncertainties are displayed as
vertical bars.}
\label{fig:cpT_centrality}
\end{figure}

The resulting $C_{p_T}$ values as a function of centrality are presented
in Fig.~\ref{fig:cpT_centrality}. The three definitions agree over most of
the centrality range. In particular, RefMult and RefMult2 are consistent
with one another within statistical uncertainties throughout, which is the
comparison that bears directly on auto-correlation: the estimator built
from the analysed particles and the estimator built from a disjoint
pseudorapidity interval return the same correlator, so the bias, if
present, is smaller than the statistical precision of this sample. A
modest difference is nonetheless visible in mid-central collisions, where
both multiplicity-based estimators depart slightly from the
impact-parameter result.

 This deviation is not an auto-correlation
effect, which would separate the two multiplicity estimators, but the
finite resolution common to any multiplicity-based centrality selection:
a given reference-multiplicity interval collects a range of impact
parameters, and the residual volume fluctuation this admits is shared by
both definitions and absent from the geometric one. Its influence is
largest where $C_{p_T}$ varies most rapidly with centrality, which is
where it is seen. The effect is small compared with the centrality
dependence of the correlator itself and does not alter the shape of the centrality dependence, but we note it
explicitly rather than describing the three estimators as equivalent.

We do not attempt to correct for this residual. The comparison that exposes
it is one between a multiplicity-based selection and the impact parameter,
and the impact parameter is not an experimental observable: removing the
effect from a measurement would require the mapping between impact parameter
and reference multiplicity to be supplied by a model, so the procedure would
replace a bounded experimental systematic with a model-dependent correction
whose own uncertainty is harder to characterise. The residual is in any case
not resolved in the most central class on
which the energy dependence of the measurement rests. The appropriate
treatment is therefore to carry it as a systematic uncertainty on the
centrality definition, which is what the comparison presented here
quantifies.

At these energies, higher-order observables such as proton cumulants
are known to be considerably more sensitive to auto-correlations and to
the limited resolution of multiplicity-based centrality
estimators~\cite{Wang_2025, Luo:2014rea, Chatterjee:2019fey}. The comparison presented here shows that $C_{p_T}$, a second-order observable constructed from all charged particles rather than from a single species is less sensitive to this effect.

The centrality bin-width correction (CBWC)~\cite{Luo:2014rea} has been applied to the data and shown to have a negligible effect on $C_{p_T}$~\cite{STAR:2026vjv}.

Taken together, the agreement between an estimator that shares its particles with the analysis acceptance and one that does not, the proximity of both to a purely geometric selection, and the insensitivity to the bin width correction establish that $C_{p_T}$ is robust against the centrality selection procedure.

\section{Robustness of $C_{p_T}$ against detector efficiency}
\label{sec:eff}

A fundamental requirement for fluctuation and correlation observables is
their robustness against finite detector efficiency, or the availability
of well-defined procedures to correct for such
effects~\cite{Luo:2013bmi,Luo:2011ts,Bzdak:2012an}. In heavy-ion
experiments, tracking efficiencies are typically below unity and depend
non-trivially on kinematic variables such as transverse momentum and
pseudorapidity~\cite{Anderson:2003ur,Llope:2012zz}. Since efficiency
effects alter event-by-event multiplicities and can distort particle
distributions, they may introduce biases in correlation measurements if
not properly controlled.

To quantify the sensitivity of $C_{p_T}$ to such effects, a closure test
is performed using events generated with the UrQMD model. The same event
sample is analysed at both the truth level and after applying a track
filter ($\epsilon(p_T,\eta)$) that emulates the tracking performance of the STAR detector,
enabling a direct comparison of the resulting observables under controlled
conditions.

\begin{figure}[htbp]
\centering
\includegraphics[width=\linewidth]{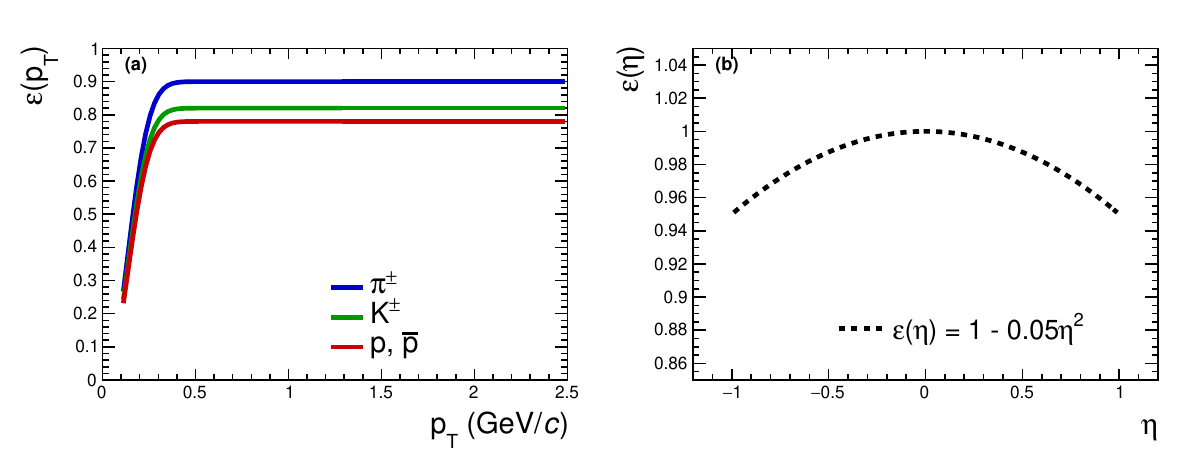}
\caption{Parameterisation of the efficiency model used in the closure
test. (a) Transverse momentum dependence of the tracking efficiency for
different particle species at midrapidity. (b) Pseudorapidity-dependent
efficiency modulation, applied uniformly to all tracks irrespective of
species.}
\label{fig:eff_model}
\end{figure}

The efficiency is implemented on a track-by-track basis using a factorised
parameterisation in transverse momentum and pseudorapidity,
\begin{equation}
\epsilon(p_T,\eta) = \epsilon(p_T)\cdot \epsilon(\eta).
\end{equation}
The transverse momentum dependence is described by a smooth turn-on
function,
\begin{equation}
\epsilon(p_T) = \epsilon_0 \left(1 - \exp\left[-\left(\frac{p_T}{p_0}\right)^{\alpha}\right]\right),
\end{equation}
with $p_0 = 0.18$~GeV/$c$ and $\alpha = 2.2$, and asymptotic efficiencies
$\epsilon_0 = 0.90$, $0.82$ and $0.78$ for pions, kaons and protons
respectively, chosen to reproduce the tracking and particle-identification
performance reported for the STAR Time Projection Chamber and
Time-of-Flight system~\cite{Anderson:2003ur,Llope:2012zz}. The pseudorapidity dependence,
\begin{equation}
\epsilon(\eta) = 1 - a \eta^2, \qquad a = 0.05,
\end{equation}
accounts for the mild acceptance variation across the detector. It is a geometric effect and is therefore common to all species: the same $\epsilon(\eta)$ multiplies every track regardless of its identity, and
all species dependence of the filter resides in $\epsilon(p_T)$. The resulting parameterisation is shown in Fig.~\ref{fig:eff_model}.

At the lower edge of the analysed interval, $p_T = 0.2$~GeV/$c$, the efficiency has reached only $72\%$ of its asymptotic value, rising to $95\%$ at
$0.3$~GeV/$c$ and saturating above $0.4$~GeV/$c$; the absolute single-track efficiency at the $p_T$ threshold is $0.65$ for pions and $0.56$ for protons. The test therefore includes a genuine momentum dependence of the bulk of the produced medium.

For each particle the corresponding efficiency is evaluated and a Bernoulli trial is performed to determine whether the track is accepted or rejected, independently of every other particle in the event.

The observables are recalculated on the efficiency-filtered particle
sample using identical kinematic selections. Collision centrality is
specified throughout by the number of participant nucleons $N_{\rm part}$
taken from the model, a geometric quantity that is by construction
unaffected by the track filter. The truth and reconstructed samples are
therefore compared event by event at fixed collision geometry, with no
possibility of migration between classes, and any residual deviation can
be attributed to the response of the correlator to particle loss alone.

In an experiment, finite efficiency enters twice: directly, through the particles entering the correlator, and indirectly, through the reconstructed multiplicity used to assign centrality. The present test
isolates the first. The second is an effect on the resolution of the centrality estimator, and is bounded by the comparison of Sec.~\ref{sec:cent}, where multiplicity-based selections were shown to
reproduce a purely geometric one within statistical uncertainties.
Treating both at once would confound two independent sources of bias and
make neither interpretable.

\begin{figure}[htbp]
\centering
\includegraphics[width=\linewidth]{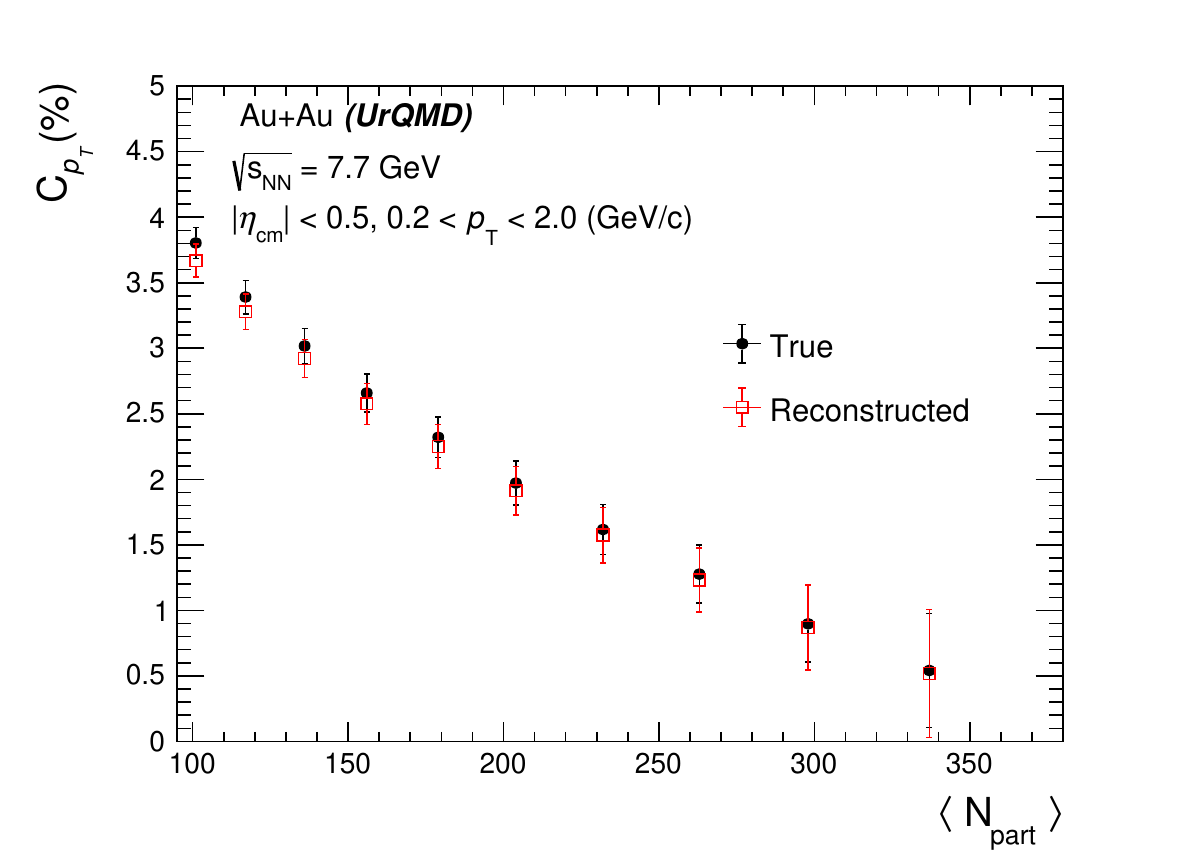}
\caption{Comparison of $C_{p_T}$ as a function of the number of
participant nucleons $N_{\rm part}$ for truth-level (solid symbols) and
efficiency-filtered (hollow symbols) samples in Au+Au collisions at
$\sqrt{s_{\mathrm{NN}}}=7.7$~GeV. Statistical uncertainties are shown as
vertical bars.}
\label{fig:cpT_eff}
\end{figure}

Figure~\ref{fig:cpT_eff} shows the centrality dependence of $C_{p_T}$ for
both the truth and the efficiency-filtered samples. The results are
consistent within statistical uncertainties across the full centrality
range, indicating that the observable is insensitive to particle loss at
the level emulated here.

This behaviour follows directly from the way the
correlator is constructed. Writing $x_i = p_{T,i} - \langle\langle p_T
\rangle\rangle$, the per-event estimator
\begin{equation}
C_{\rm ev} = \frac{\sum_{i \neq j} x_i x_j}{N(N-1)}
\end{equation}
is normalised by the number of pairs rather than by any extensive
quantity. If each particle is retained independently with probability
$\epsilon_i$, then for a fixed underlying event the expectation values of
numerator and denominator over the filter are
\begin{equation}
\Big\langle \sum_{i \neq j \in A} x_i x_j \Big\rangle = \sum_{i \neq j} \epsilon_i \epsilon_j \, x_i x_j,
\qquad
\big\langle n(n-1) \big\rangle = \sum_{i \neq j} \epsilon_i \epsilon_j ,
\end{equation}
where $A$ denotes the accepted subset and $n = |A|$. The measured
correlator is therefore the pair covariance weighted by $\epsilon_i
\epsilon_j$, and two cases are distinguished.

For an efficiency that does not depend on transverse momentum, $\epsilon_i
= \epsilon_0$, the weights are common to every pair and cancel identically
between numerator and denominator, for any value of $\epsilon_0$. The
cancellation in this case is exact rather than approximate: under uniform thinning the accepted set is a
simple random subsample of the parent event, and $C_{\rm ev}$ is a
$U$-statistic~\cite{hoeffding1948class} over that subsample, hence an unbiased estimator of the parent
pair average for every retained multiplicity $n\geq2$. It holds event by
event and not merely on average over events. This statement
concerns the pair average alone; the normalisation by
$\langle\langle p_{\rm T}\rangle\rangle^{2}$ plays no part in it, since a
uniformly thinned sample leaves the transverse-momentum distribution, and
therefore the mean, unchanged.

For an efficiency that does depend on transverse momentum no such identity
exists, and neither the covariance nor the mean is individually invariant.
What protects $C_{p_T}$ in this regime is that both are evaluated on the
same re-weighted transverse-momentum distribution, so that the leading
effect of the re-weighting --- the displacement of the mean --- divides out
in the ratio. What does not divide out is the change in the shape of the
distribution relative to its own mean, and this is the residual sensitivity
the closure test of this section is designed to expose. That the residual
is small is an empirical statement about the efficiency rather than an
identity: over the range $0.2 < p_T < 2.0$~GeV/$c$ used here the
parameterisation of Fig.~\ref{fig:eff_model} is close to its plateau across
most of the yield, so the re-weighting is mild, and the agreement seen in
Fig.~\ref{fig:cpT_eff} is the quantitative expression of that statement.
The species dependence of $\epsilon(p_T)$ enters only through this same
mechanism, as a mild change in the relative pion, kaon and proton weights
within the accepted sample.

$C_{p_T}$ is insensitive to how many particles are lost, and sensitive only to whether
the loss is momentum dependent --- and then only through the variation of
the efficiency across the analysed interval, not through its magnitude.

This places $C_{p_T}$ in a different position from cumulant-based
observables, where efficiency enters through factorial moments and must be
removed by an explicit binomial unfolding~\cite{Luo:2011ts,Bzdak:2012an}. No such correction procedure is required here. The correlator may therefore be compared directly across detector configurations, running periods and collision energies without a common efficiency being assumed --- a property of particular value for the Beam Energy Scan and the fixed-target programme, where tracking performance and acceptance vary substantially between data sets.

\section{Robustness of $C_{p_T}$ against acceptance effects}
\label{sec:acceptance}

A comparison of measurements performed in collider and fixed-target configurations at overlapping centre-of-mass energies provides a test of whether an observable depends on the accelerator running mode. STAR has shown that $C_{p_T}$ is consistent between these configurations when analysed in equivalent kinematic regions~\cite{STAR:2026vjv}. Such consistency is important for combining measurements across a broad range of baryon chemical potentials and is relevant beyond STAR. Fluctuation observables are being studied at high baryon densities in fixed-target experiments such as HADES at SIS18~\cite{HADES:2009aat}, and will be investigated by CBM at FAIR~\cite{Friman:2011zz,Ablyazimov:2017guv} and BM@N at NICA~\cite{Kekelidze:2016hhw,Kapishin:2020oah}. MPD at NICA~\cite{MPD:2022qhn}, designed for collider operation, has also considered a fixed-target mode~\cite{Parfenov:2024zwl}; an overview of these programmes is given in Ref.~\cite{Senger:2021tlg}. Fluctuation observables have already been measured in fixed-target geometry in this regime~\cite{HADES:2020wpc}. Understanding how the change in running mode affects particle selection is therefore important for comparisons across these experiments.

The transverse momentum of a particle is invariant under a longitudinal Lorentz boost. Consequently, if the same events and particles were selected in both configurations, $C_{p_T}$ would be identical. The non-trivial aspect of this comparison arises from the acceptance. In fixed-target geometry, particle selection is imposed in the laboratory pseudorapidity, and unlike rapidity, pseudorapidity is not additive under longitudinal boosts. A cut in $\eta_{\rm lab}$ therefore does not map exactly onto the corresponding pseudorapidity interval in the centre-of-mass frame, with the mapping depending on particle mass and transverse momentum. The two configurations can consequently select different particles from the same underlying events, and the resulting sensitivity of the correlator to these differences is what is examined here.

UrQMD events are generated in the nucleon--nucleon centre-of-mass frame, which coincides with the laboratory frame for symmetric collider running. To emulate fixed-target kinematics, all particles are boosted to the laboratory frame. Since rapidity is additive under longitudinal Lorentz boosts, the transformation is implemented as
\begin{equation}
y_{\mathrm{lab}} = y_{\mathrm{cm}} + y_{\mathrm{shift}},
\end{equation}
where $y_{\rm shift}$ is the rapidity of the centre-of-mass frame relative to the laboratory frame,

$$
    |y_{\rm shift}| =
    \operatorname{arccosh}\!\left(\frac{\sqrt{s_{NN}}}{2m_N}\right),
$$

equal to $2.09$ at $\sqrt{s_{NN}} = 7.7$~GeV. The sign is determined by the beam-direction convention, with the target at rest in the laboratory frame.

After the boost, the laboratory pseudorapidity is recalculated from the particle rapidity and transverse momentum according to
\begin{equation}
\eta_{\mathrm{lab}} = \frac{1}{2} \ln \left(
\frac{\sqrt{m_T^2 \cosh^2 y_{\mathrm{lab}} - m^2} + m_T \sinh y_{\mathrm{lab}}}
{\sqrt{m_T^2 \cosh^2 y_{\mathrm{lab}} - m^2} - m_T \sinh y_{\mathrm{lab}}}
\right),
\label{eq:eta}
\end{equation}
where $m_T = \sqrt{m^2 + p_T^2}$ is the transverse mass. This transformation introduces a mass dependence into the mapping between rapidity and pseudorapidity: particles with the same rapidity and transverse momentum but different masses acquire different pseudorapidities and can therefore be accepted or rejected differently by the same pseudorapidity cut.

\begin{figure}[htbp]
\centering
\includegraphics[width=\linewidth]{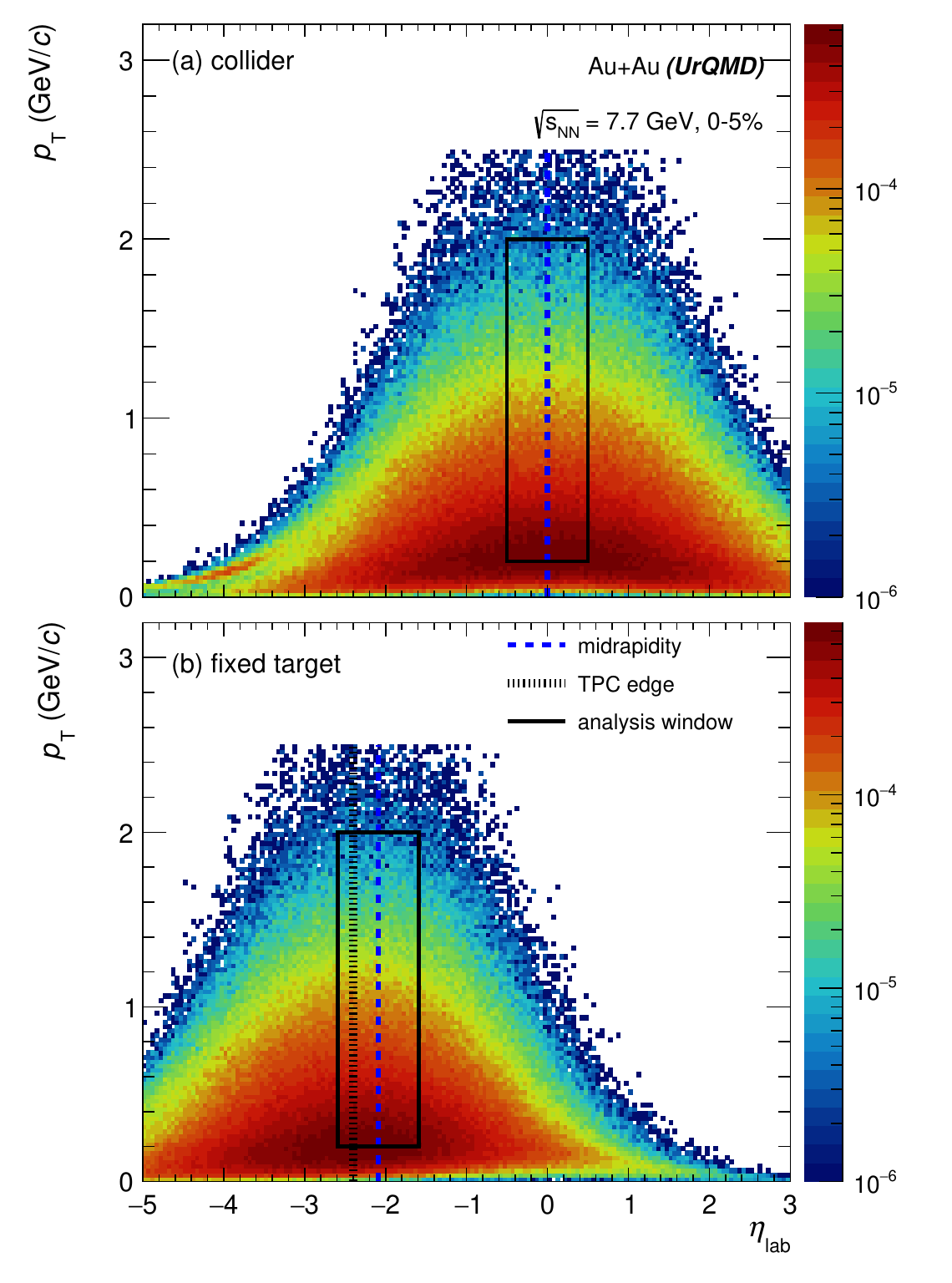}
\caption{Distribution of charged particles in transverse momentum and
laboratory pseudorapidity in $0$--$5\%$ central Au+Au collisions at
$\sqrt{s_{\mathrm{NN}}}=7.7$~GeV in UrQMD. (a) Collider configuration.
(b) Fixed-target configuration, obtained by boosting the same events by
$y_{\rm shift}=-2.09$. The dashed lines mark midrapidity, the boxes the
analysis window $|\eta_{\rm cm}|<0.5$, $0.2<p_T<2.0$~GeV/$c$, and the dotted
line in (b) the edge of the TPC at $\eta_{\rm lab}=-2.4$. Each panel is
normalised to unit integral.}
\label{fig:pteta}
\end{figure}

The fixed-target acceptance is then applied. The STAR TPC, together with the iTPC upgrade, covers $-2.4<\eta_{\rm lab}<0$. The analysis uses $|\eta_{\rm cm}|<0.5$ and $0.2<p_T<2.0$~GeV/$c$, where the shifted pseudorapidity coordinate is defined as $\eta_{\rm cm}=\eta_{\rm lab}-\eta_{\rm mid}$~\cite{STAR:2026vjv}. At $\sqrt{s_{NN}}=7.7$~GeV, $\eta_{\rm mid}=-2.09$, so the nominal analysis window corresponds to $-2.59<\eta_{\rm lab}<-1.59$, extending beyond the detector's lower pseudorapidity boundary. Figure~\ref{fig:pteta} shows the charged-particle distributions in $p_T$ and $\eta_{\rm lab}$ for central UrQMD events in the collider and fixed-target configurations, together with their respective analysis windows. In the UrQMD sample, the part of the nominal fixed-target window below $\eta_{\rm lab}=-2.4$ contains $20\%$ of the charged particles that would otherwise fall within that window. After the detector acceptance is imposed, the fixed-target selection contains $8\%$ fewer charged particles than the collider selection. Moreover, only $66\%$ of the particles accepted in the collider configuration are also accepted in the fixed-target configuration. The acceptance loss is therefore not equivalent to a uniform reduction in multiplicity: the mass dependence of Eq.~(\ref{eq:eta}) shifts heavier particles at low $p_T$ towards more negative $\eta_{\rm lab}$, resulting in species- and momentum-dependent losses. The corresponding truncation at the detector boundary is also visible in the measured $p_T$--$\eta_{\rm lab}$ distributions for the $7.7$~GeV collider and fixed-target samples in Fig.~1 of Ref.~\cite{STAR:2026vjv}.

\begin{figure}[htbp]
\centering
\includegraphics[width=\linewidth]{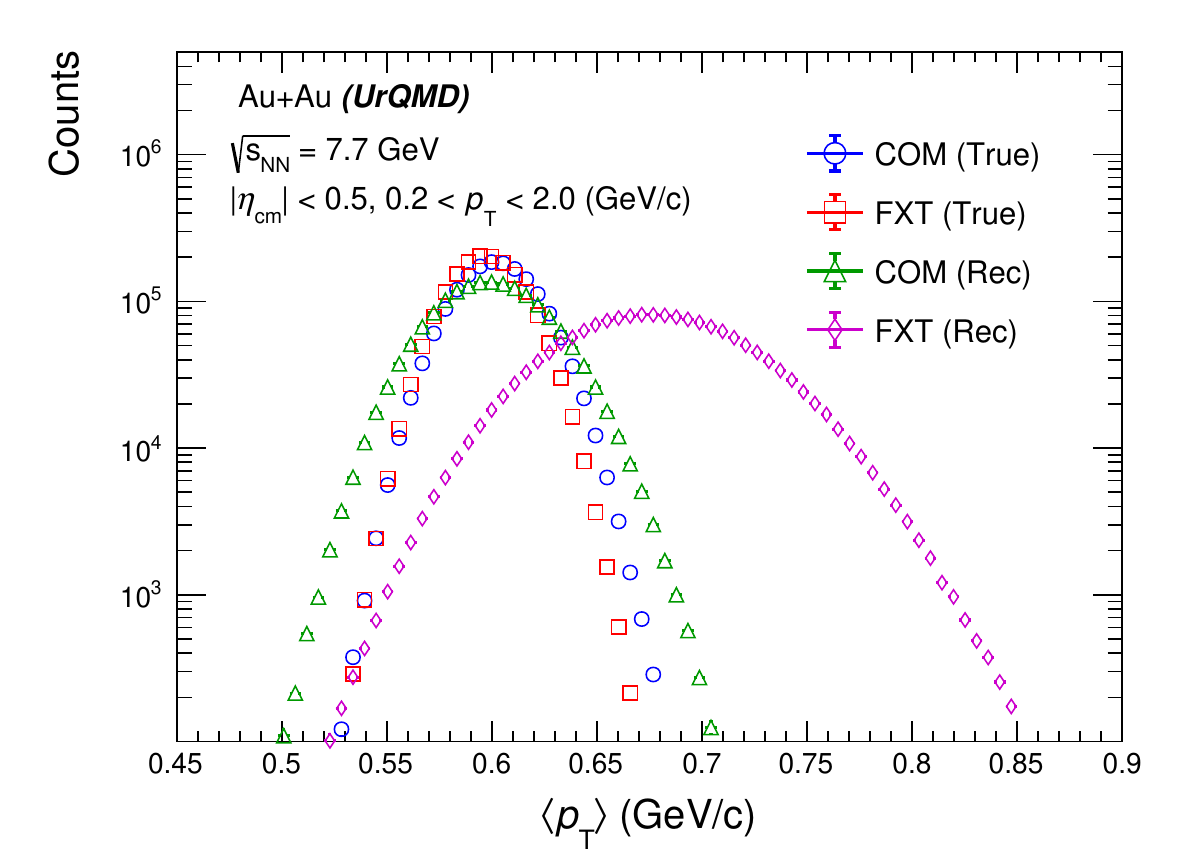}
\caption{Mean transverse momentum $\langle p_{\mathrm{T}}\rangle$ distributions in  Au+Au collisions at $\sqrt{s_{\mathrm{NN}}}=7.7$~GeV, comparing collider (COM) and fixed-target (FXT) configurations at the generator (true) and reconstructed levels.}
\label{fig:Mean_pt}
\end{figure}

The effect of this loss on a single-particle observable is seen in the mean
transverse momentum distributions for collider and fixed-target
configurations at both the generator (``true'') and reconstructed levels,
shown in Fig.~\ref{fig:Mean_pt}. The reconstructed-level results exhibit
clear modifications relative to the true distributions, reflecting the
combined effect of tracking efficiency and the loss of acceptance
described above. A deviation of the same character and comparable
magnitude between the collider and fixed-target mean $p_T$ is present in
the STAR data themselves~\cite{STAR:2026vjv}, so the detector response
implemented here reproduces the single-particle discrepancy seen in the
data, and the comparison below tests the correlator against a realistic
distortion.

\begin{figure}[htbp]
\centering
\includegraphics[width=\linewidth]{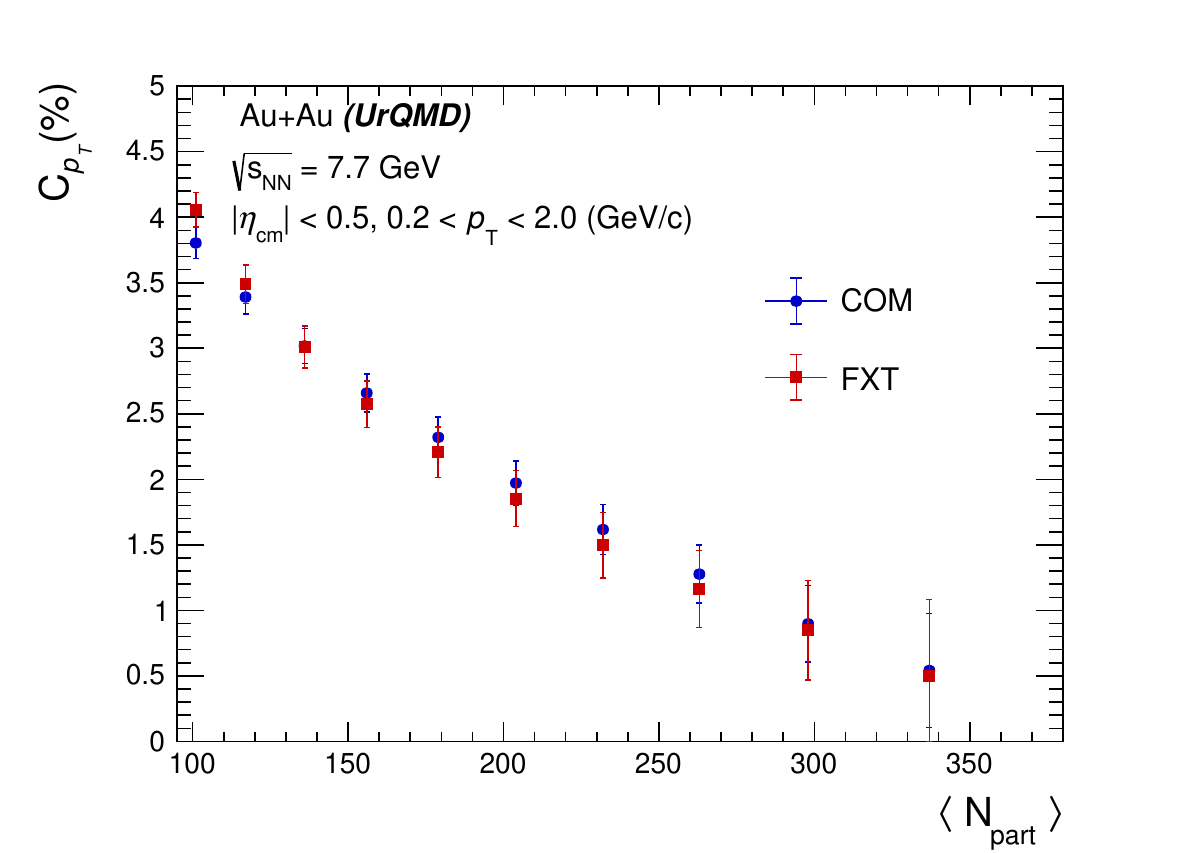}
\caption{Comparison of $C_{p_T}$ as a function of centrality for COM mode (blue solid symbols) and FXT mode (red solid symbols) samples in Au+Au collisions at $\sqrt{s_{\mathrm{NN}}}=7.7$~GeV. Statistical uncertainties are shown as vertical bars.}
\label{fig:cpT_frame_eff}
\end{figure}

The two-particle correlator $C_{p_T}$ is shown in Fig.~\ref{fig:cpT_frame_eff}. Despite the loss of particles, it is unchanged. The values obtained in the boosted (fixed-target) and unboosted (collider) configurations agree within statistical uncertainties over most of the centrality range, both at the generator and at the reconstructed level.

The comparison therefore shows that a loss of acceptance which is plainly
visible at the level of $\langle p_T \rangle$ leaves the scaled correlator
unchanged. This is the expected behaviour: the loss of acceptance and the
tracking efficiency both act by re-weighting the single-particle
transverse momentum distribution, and $C_{p_T}$ is constructed as a ratio
in which a common re-weighting of the mean and of the pair covariance
cancels, as set out in Sec.~\ref{sec:eff}.

One limitation should be stated. The fixed-target sample analysed here is
obtained by boosting events generated in the centre-of-mass frame, so the
test isolates the effect of the frame-dependent acceptance while leaving
the underlying event generation untouched. It does not probe any residual
frame dependence of the transport calculation itself, which enters through
the frame in which the collision criterion and the time ordering of
interactions are defined. A direct comparison against events generated in
the target frame would close that gap, and we note it as the natural
extension of the present test rather than as a result of it.

$C_{p_T}$ may therefore be compared between collider and fixed-target
data without an acceptance correction.

\section{Decomposition of the correlator}
\label{sec:decomp}

The preceding sections establish that $C_{p_T}$ is insensitive to the choice
of multiplicity-based centrality estimator, to the tracking efficiency, and to the acceptance
in which the measurement is made. Taken together these are the
properties that qualify an observable to enable a comparison across collision
energies. It is on this basis that the non-monotonic energy
dependence reported by STAR is established with a significance of
$5\sigma$~\cite{STAR:2026vjv}.

It has recently been argued by Reichert \textit{et al.}~\cite{Reichert:2026mbm} that the non-monotonicity
requires no critical dynamics and follows instead from the changing hadronic
composition of the final state: mesons and baryons carry different mean
transverse momenta, their relative abundance varies with collision energy,
and a mixture of two populations with different means contributes to the
correlator. The argument is developed within a schematic two-component
parameterisation rather than a transport calculation, but it takes its
multiplicity yields from the UrQMD final state and can therefore be tested
directly within the framework used throughout this work. We first decompose
the correlator into its two statistical components and then examine how
meson--baryon mixing contributes to each of them.

The decomposition is shown at $7.7$~GeV. Meson--baryon mixing, by contrast,
has to be followed across the range of energies over which the composition of
the accepted sample changes. We therefore generated $10^{4}$ minimum-bias
events at each of sixteen energies, $\sqrt{s_{NN}} = 2.0$, $2.5$, $3.0$,
$3.2$, $3.5$, $3.9$, $4.5$, $5.2$, $7.7$, $11.5$, $14.6$, $19.6$, $27.0$,
$39.0$, $62.4$ and $200$~GeV, covering the STAR fixed-target points, the
collider points of the beam-energy scan, and the full range over which the
two-component model of Reichert \textit{et al.} is evaluated, and analysed each of them
in the same acceptance and centrality class as the measurement. The meson and
baryon fractions and their event-by-event variance, tabulated in
Appendix~\ref{app:model}, are inclusive properties of the accepted sample and
converge rapidly: $10^{4}$ minimum-bias events per energy, of which
approximately $500$ fall in the $0$--$5\%$ class, determine them to a few
percent. The correlator itself, being the residual of a cancellation at the
percent level, would require samples several orders of magnitude larger, and
we do not quote it from these samples.

Meson--baryon mixing matters only in the middle of this range. Within the
acceptance of the measurement the baryon fraction of the accepted charged
particles falls monotonically from $f_B = 0.64$ at $3.0$~GeV, through
$f_B = 0.51$ at $3.5$~GeV, to $f_B = 0.09$ at $14.6$~GeV and $f_B = 0.07$ at
$19.6$~GeV. The mixing contribution derived below scales as $f_B(1-f_B)$ and
is therefore already an order of magnitude below its maximum by $15$~GeV.
Above that energy the accepted sample is almost purely mesonic; the points
are retained because the two-component calculation is quoted over the full
range, and for the correlator itself the published UrQMD baseline of
Ref.~\cite{STAR:2019clv} applies.

The pair sum in Eq.~(\ref{eq:corr}) can be rewritten exactly. For a single event,
\begin{equation}
C_{m1,k}=\langle p_{\rm T}\rangle_{k}^{2}-\frac{s^{2}_{p_T,k}}{N_k-1},
\end{equation}
where $s^{2}_{p_T,k}$ is the transverse-momentum variance of the accepted particles in that event. Averaging over events and inserting this into Eq.~(\ref{eq:cm1}) gives
\begin{equation}
\langle \Delta p_{T,i}\,\Delta p_{T,j}\rangle
=\mathrm{Var}\!\left(\langle p_{\rm T}\rangle_{k}\right)
-\left\langle \frac{s^{2}_{p_T,k}}{N_k-1}\right\rangle .
\label{eq:decomp}
\end{equation}
The first term is the event-by-event fluctuation of the mean transverse momentum; the second is the mean in-event transverse-momentum variance. Both are positive, and both follow from the same two distributions used for Eq.~(\ref{eq:cm1}).

\begin{figure}[htbp]
\centering
\includegraphics[width=\linewidth]{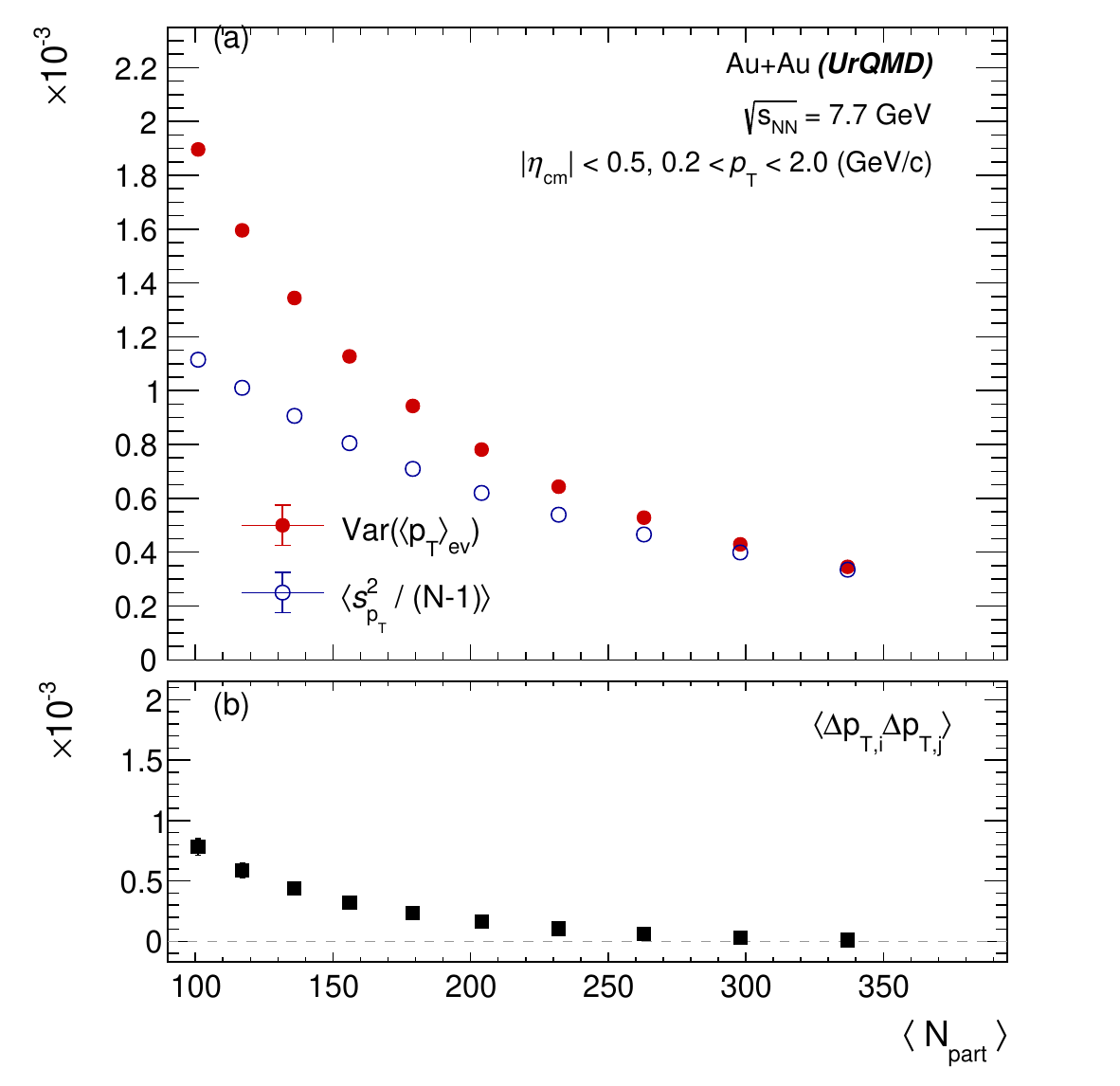}
\caption{(a) The two terms of Eq.~(\ref{eq:decomp}) as a function of centrality in Au+Au collisions at $\sqrt{s_{\mathrm{NN}}}=7.7$~GeV in UrQMD: the event-by-event variance of the mean transverse momentum (red solid circles) and the mean in-event transverse-momentum variance (blue open circles). (b) The difference of the two, $\langle \Delta p_{T,i}\,\Delta p_{T,j}\rangle$. The dashed line marks zero. Statistical uncertainties are obtained with the bootstrap method.}
\label{fig:decomp}
\end{figure}

Figure~\ref{fig:decomp} shows the two terms separately in UrQMD at $\sqrt{s_{\mathrm{NN}}}=7.7$~GeV. They are non-zero and nearly equal: in the $0$--$5\%$ class they are $3.46\times10^{-4}$ and $3.36\times10^{-4}$, so the correlator that survives between them is a $3\%$ residual. The cancellation weakens toward peripheral collisions, where the residual grows to roughly $40\%$ of either term.

This has a direct consequence for the interpretation of the observable. Any mechanism that shifts either term at the few-percent level changes the correlator by an amount comparable to its entire value, and meson--baryon mixing has to be assessed with this in mind.

\subsection{Sensitivity to the composition-fluctuation spectrum}
\label{sec:composition}

Consider $N$ accepted particles drawn from a two-component sample: a fraction $f$ of baryons with mean transverse momentum $\bar{p}_B$ and variance $\sigma_B^{2}$, and a remaining fraction $1-f$ of mesons with $\bar{p}_M$ and $\sigma_M^{2}$. Write $\Delta\bar{p}=\bar{p}_B-\bar{p}_M$. Here $f$ is the baryon fraction, written $f_B$ elsewhere in this work and in the tables of Appendix~\ref{app:model}; the combination $f(1-f)$ below is symmetric under the exchange of the two species. The single-particle transverse-momentum distribution of the mixture has variance
\begin{equation}
\sigma^{2}_{\mathrm{mix}}=f\sigma_B^{2}+(1-f)\sigma_M^{2}+f(1-f)\Delta\bar{p}^{2},
\label{eq:mixvar}
\end{equation}
where the last term is the variance of the two-valued distribution of species means and is the only place the composition enters.

Suppose that the baryon-meson assignments of each particle is independent, so that the species counts fluctuate binomially and the accepted transverse momenta are independent, identically distributed samples of the mixture. The first term of Eq.~(\ref{eq:decomp}) is then the variance of a sample mean of $N$ such draws, $\sigma^{2}_{\mathrm{mix}}/N$. The second follows from $\langle s^{2}_{p_T,k}\rangle=\frac{N-1}{N}\,\sigma^{2}_{\mathrm{mix}}$, which gives $\langle s^{2}_{p_T,k}/(N-1)\rangle=\sigma^{2}_{\mathrm{mix}}/N$ as well. The two terms cancel identically and the correlator vanishes, as it must for statistically independent emission --- including the composition term, whose contribution to the in-event variance is exactly compensated by the contribution that event-by-event fluctuations of the composition make to the variance of the event mean.

Suppose instead that the species counts are held fixed at $N_B=fN$ in every event. The event mean then has variance $[f\sigma_B^{2}+(1-f)\sigma_M^{2}]/N$, the composition term being absent because the composition no longer fluctuates. The in-event variance is unaffected, since a sample of fixed composition still contains both species; evaluating it for fixed $N_B$ and $N_M$ gives $\langle s^{2}_{p_T,k}/(N_k-1)\rangle = [f\sigma_B^{2}+(1-f)\sigma_M^{2}]/N + f(1-f)\Delta\bar{p}^{2}/(N-1)$. The difference is
\begin{equation}
\langle \Delta p_{T,i}\,\Delta p_{T,j}\rangle=-\,\frac{f(1-f)\,\Delta\bar{p}^{2}}{N-1},
\label{eq:frozen}
\end{equation}
which is negative and largest in magnitude at $f=1/2$, that is, where the meson and baryon yields are comparable. Equation~(\ref{eq:frozen}) is identical to the expression obtained by Reichert \textit{et al.}, whose $N_M N_B/[N^{2}(N-1)]$ is $f(1-f)/(N-1)$.

The composition contribution is therefore not determined by the mean yields alone. It is set by the event-by-event composition-fluctuation spectrum, and it interpolates continuously between zero and Eq.~(\ref{eq:frozen}) as that spectrum is varied between the binomial and the frozen limit. In heavy-ion collisions the spectrum is sub-binomial, since baryon number is conserved, so the true contribution lies between these limits; which is an output of the dynamics rather than an input.

Within a fixed centrality class and the same
acceptance applied to the correlator, we extract the variance of the event
baryon fraction $f=N_B/N$ directly from the transport final state and compare
it with the binomial value $f(1-f)/N$. Two precautions are necessary: the
measurement must be made within a narrow centrality class, since evaluated
over a minimum-bias sample the apparent composition variance is inflated by
the mixing of collision geometries and exceeds the binomial value by a large
factor that carries no information about the mixing; and the
acceptance must match that of the correlator.

The ratio $\mathrm{Var}(f)/[f(1-f)/N]$ is extracted at all sixteen energies
of the scan and is sub-binomial for every one of them. Across the ten energies
between $3.0$ and $19.6$~GeV, where meson--baryon mixing is relevant, it
lies between $0.81$ and $0.97$, with a mean of $0.89\pm0.02$ and no
significant energy dependence: a constant describes the ten values with
$\chi^{2}/\mathrm{ndf}=8.4/9$. Over the full scan, $2.5$--$200$~GeV, the mean
is $0.92\pm0.02$; we quote the lower value because it is the one that applies
where the mixing contributes. The composition-fluctuation spectrum is therefore sub-binomial, as baryon-number conservation requires,
but only mildly so. It lies close to the independent-emission limit, in which
the composition contribution cancels identically, and far from the frozen
limit of Eq.~(\ref{eq:frozen}), in which it is maximal. Since the net
contribution is proportional to $1-\mathrm{Var}(f)/[f(1-f)/N]$, the
mixing term retained by Reichert \textit{et al.} is reduced to
$11\%$ of its frozen-limit value.

We note that Reichert \textit{et al.} anticipate this effect. Among the
contributions listed there as neglected are fluctuations of $N_{ch}$, $N_M$
and $N_B$ which can be correlated, with the observation that one would expect
$N_M/N_B$ to fluctuate other than randomly and that such an effect reduces
the overall magnitude of the covariance, bringing it closer to zero. The
measurement presented here quantifies the magnitude of that reduction.

Reichert \textit{et al.} evaluate the frozen-composition limit of Eq.~(\ref{eq:frozen}) and combines it with a partial-momentum-conservation contribution of the opposite sign whose strength is a free parameter of that calculation. Each contribution is individually comparable to, or larger than, the measured correlator, so their sum is sensitive to the relative normalisation of the two. A transport calculation fixes both self-consistently: it contains the same change of hadronic composition, together with exact momentum conservation, resonance production and decay, and the composition-fluctuation spectrum generated by the collision dynamics. Section~\ref{sec:excitation} carries out that comparison.

\subsection{Consequences for the excitation function}
\label{sec:excitation}

\begin{figure}[htbp]
\centering
\includegraphics[width=\linewidth]{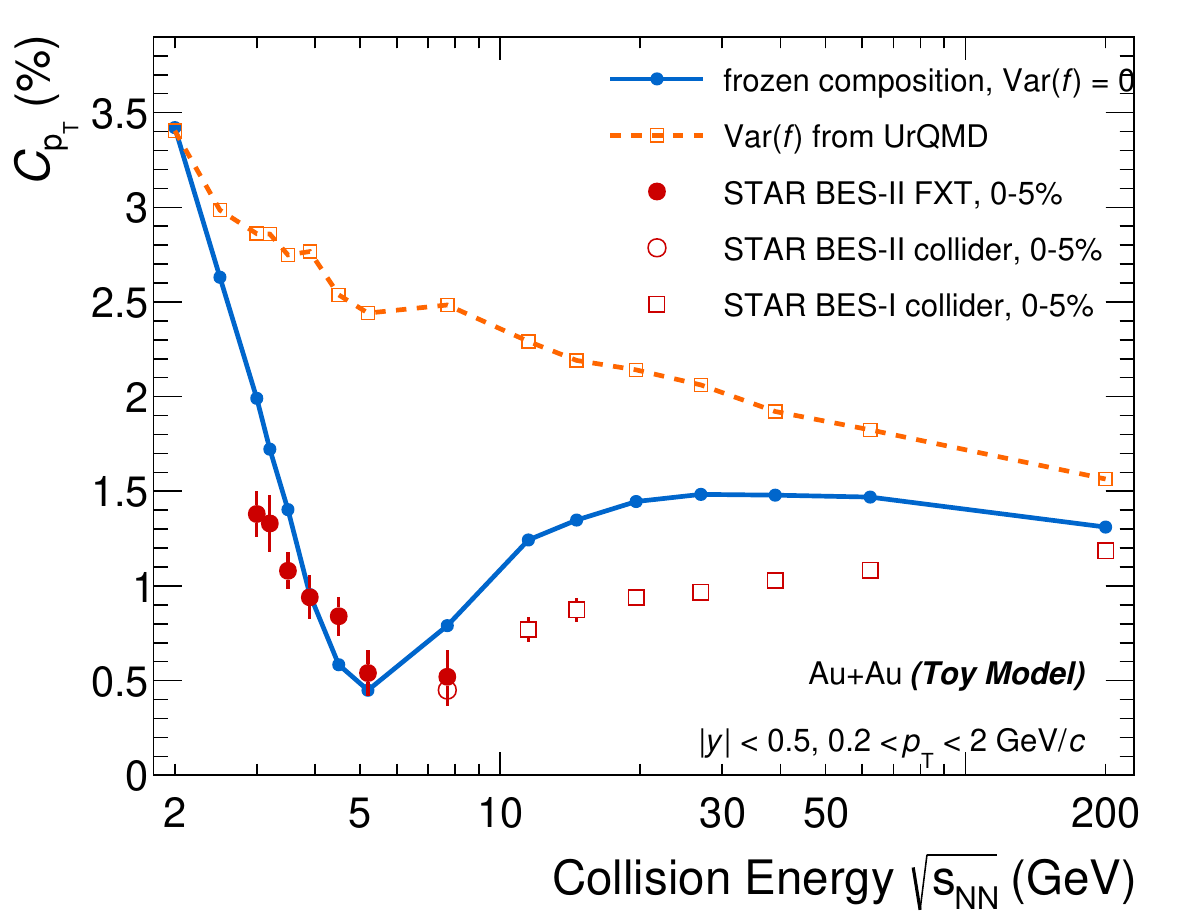}
\caption{The two-component meson--baryon contribution to the correlator,
expressed as $C_{p_T}$, evaluated in the acceptance $|y|<0.5$,
$0.2<p_T<2$~GeV/$c$ of Reichert \textit{et al.}~\cite{Reichert:2026mbm} with their effective temperatures
and their partial-pairing treatment of momentum conservation. Solid blue circles:
the species counts held fixed event to event, $\mathrm{Var}(f)=0$,
reproducing Reichert \textit{et al.}. Open orange squares: the identical calculation with the composition fluctuation measured in the UrQMD final state. Solid red circles are the
measured $0$--$5\%$ central values~\cite{STAR:2026vjv}, filled for the
fixed-target and open red circle and red squares for the collider datasets, with statistical and systematic uncertainties added in quadrature. Model uncertainties are the
standard error over ten independent evaluations of $4\times10^{5}$ events.}
\label{fig:mixing}
\end{figure}

The two ingredients are now in place: the frozen-composition expression of
Eq.~(\ref{eq:frozen}), which is the limit evaluated by Reichert \textit{et al.}, and the composition-fluctuation spectrum the
transport calculation actually produces. Figure~\ref{fig:mixing} compares them, using the conditions adopted by Reichert \textit{et al.}, namely the acceptance ($|y|<0.5$, $0.2<p_T<2$~GeV/$c$), their two-component final state of charged pions and protons, their effective temperatures ($T^{p}_{\rm eff}=280$~MeV,
$T^{\pi}_{\rm eff}=130$~MeV), and their implementation of momentum conservation
as pairwise transverse-momentum balancing with probability $Z/A$ for protons
and $2/3$ for pions. Under these conditions we reproduce their calculation quantitatively across the full range
$\sqrt{s_{NN}}=2$--$200$~GeV, in each of its three momentum-conservation
scenarios; the reproduction is documented in Appendix~\ref{app:model}. That agreement (solid curve) anchors the comparison to the UrQMD variation (dashed curve) which differs in exactly one input, namely the composition fluctuation. The solid curve is non-monotonic, with a minimum between $4$ and $6$~GeV, below that of the data. Across the whole range it stays within a factor of two of the measurement.
The calculation represented by the dashed curve changes only the baryon-meson composition fluctuation, which was set to zero in the solid curve. Instead it is taken from the UrQMD final state, at the $\mathrm{Var}(f)/[f(1-f)/N]=0.89\pm0.02$ measured in Sec.~\ref{sec:composition}, temperatures, yields and pairing probability are all held fixed. The minimum disappears. The curve is monotonic from $2$ to $200$~GeV, and in the region of the measured minimum it exceeds the data by a factor of about five rather than reproducing it. Thus the central result of this study is that the minimum of the two-component calculation is not a consequence of the changing meson-to-baryon ratio, which is present in both curves and identical between them. It is rather a consequence of holding the ratio fixed event by event.

With the composition fluctuation taken from transport, meson--baryon mixing does not generate a non-monotonic excitation function at any strength of the momentum-conservation term (Appendix~\ref{app:model}).

\section{Conclusion}

Measurements of $C_{p_T}$, or observables directly related to it, span more than three orders of magnitude in collision energy, from $p+p$ collisions at the ISR~\cite{CERN-Dortmund-Heidelberg-Warsaw-Ames-Bologna:1983asn} and the LHC~\cite{ALICE:2014ccw,ALICE:2025pTmoments,ALICE:2025epj} to heavy-ion collisions at the SPS~\cite{CERES:2008wlj}, RHIC~\cite{STAR:2019clv,STAR:2026vjv}, and the LHC~\cite{ALICE:2014ccw,ALICE:2025pTmoments,ALICE:2025epj}. While the correlator exhibits comparatively weak energy dependence at high collision energies, a pronounced non-monotonic structure emerges in central Au+Au collisions at low energies, corresponding to the high-baryon-density region of the QCD phase diagram. Our results demonstrate that the observed non-monotonicity is robust against the experimental and methodological variations investigated here and cannot be explained by the meson--baryon mixing prescription examined in this work. Recent theoretical studies predict non-monotonic behaviour in temperature-fluctuation observables near the QCD phase transition and critical point~\cite{gyure2026fourdimensionalqcdequationstate}, motivating quantitative comparisons with the measured $C_{p_T}$. Further measurements in this energy regime, particularly in $p+p$ collisions and heavy-ion collisions with HADES at SIS18 and CBM at FAIR~\cite{HADES:2009aat,Friman:2011zz,Ablyazimov:2017guv}, would provide valuable complementary information. Establishing the origin of this structure will require further experimental measurements and quantitative comparisons with theoretical calculations that consistently account for critical dynamics and other contributions to the observed correlations.


\begin{acknowledgments}
We thank Fuqiang Wang and Roy Lacey for useful discussions. This work was supported by the U.S. Department of Energy, Office of Science, under Award No. DE-FG02-07ER41521. We also thank the UrQMD collaboration for making their code publicly available.

\end{acknowledgments}

\section*{Data availability}
The numerical data supporting the findings of this article were generated using the UrQMD transport model. The processed data underlying the figures are available from the corresponding author upon reasonable request. The STAR experimental results used for comparison are reported in Ref.~\cite{STAR:2026vjv}. The UrQMD source code is publicly available [\url{urqmd.org}].

\appendix

\section{The two-component model: reproduction and variations}
\label{app:model}

This appendix documents the reproduction of the calculation of
Reichert \textit{et al.}~\cite{Reichert:2026mbm} that underlies Fig.~\ref{fig:mixing}, and the two
variations kept out of the main text: the strength of the
momentum-conservation term, and the choice of acceptance. Neither changes the
conclusion.

Some of the curves shown here are negative, which $C_{p_T}$ cannot
represent. In this appendix we therefore use the linear variable
\begin{equation}
\mathcal{L}=\frac{\langle \Delta p_{T,i}\,\Delta p_{T,j}\rangle}{\langle\langle p_{\rm T}\rangle\rangle^{2}},
\label{eq:linear}
\end{equation}
which equals $C_{p_T}^{2}$ wherever the correlator is positive and remains
defined where it is not. Negative correlations are physically meaningful:
they can arise from global momentum conservation and from sub-binomial
fluctuations of the hadronic composition. $\mathcal{L}$ is the quantity
reported by the ATLAS Collaboration as $k_2$~\cite{PhysRevLett.133.252301};
relations to the other measures in use are given in
Appendix~\ref{app:measures}.

\subsection{Reproduction}

\begin{figure*}[htbp]
\centering
\includegraphics[width=0.92\textwidth]{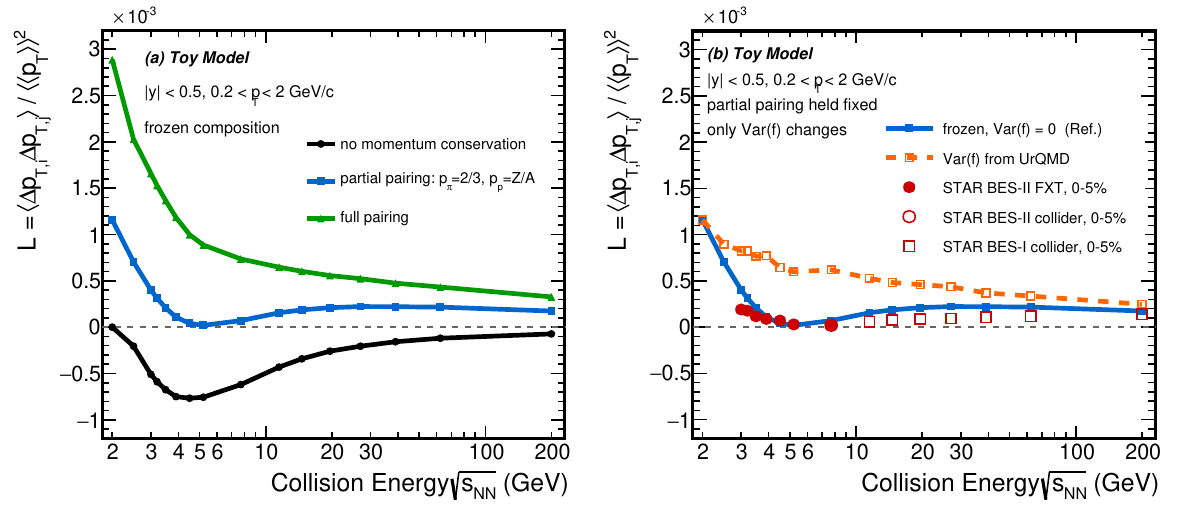}
\caption{(a) Reproduction of the three momentum-conservation scenarios of
Reichert \textit{et al.}~\cite{Reichert:2026mbm}, Fig.~1, with frozen composition in their
acceptance. (b) The partial-pairing case with the composition fluctuation
held at zero and taken from UrQMD, with the measurement~\cite{STAR:2026vjv}.}
\label{fig:repro}
\end{figure*}

Reichert \textit{et al.} implement momentum conservation by drawing
a fraction of the particles in pairs carrying equal transverse momentum, the
pairing occurring with probability $Z/A=79/197$ for protons and $2/3$ for
pions, and presents three scenarios: no pairing, partial pairing, and full
pairing. Adopting their acceptance, effective temperatures and frozen
composition, and taking the yields from the UrQMD scan of
Table~\ref{tab:inputs_y}, we obtain the values of Table~\ref{tab:repro}.
These agree with their Fig.~1 across $\sqrt{s_{NN}}=2$--$200$~GeV: the
no-conservation curve is negative with a minimum of $-7.7\times10^{-4}$ near
$4.5$~GeV, the partial-pairing curve passes close to zero near $5$~GeV, and
the full-pairing curve falls from $2.9\times10^{-3}$ at $2$~GeV to
$3.2\times10^{-4}$ at $200$~GeV. Figure~\ref{fig:repro} shows the three
scenarios together.

We note that meson--baryon mixing is inoperative at both ends of the range, for the
same reason. The composition contribution scales as $f_B(1-f_B)$, which
vanishes when either species dominates: at $2$~GeV the accepted sample is
essentially pure baryons ($f_B=0.9999$, Table~\ref{tab:inputs_eta}) and at
$200$~GeV essentially pure mesons ($f_B=0.032$). The structure can therefore
only be generated in between.

\subsection{Strength of the momentum-conservation term}

\begin{figure*}[htbp]
\centering
\includegraphics[width=0.92\textwidth]{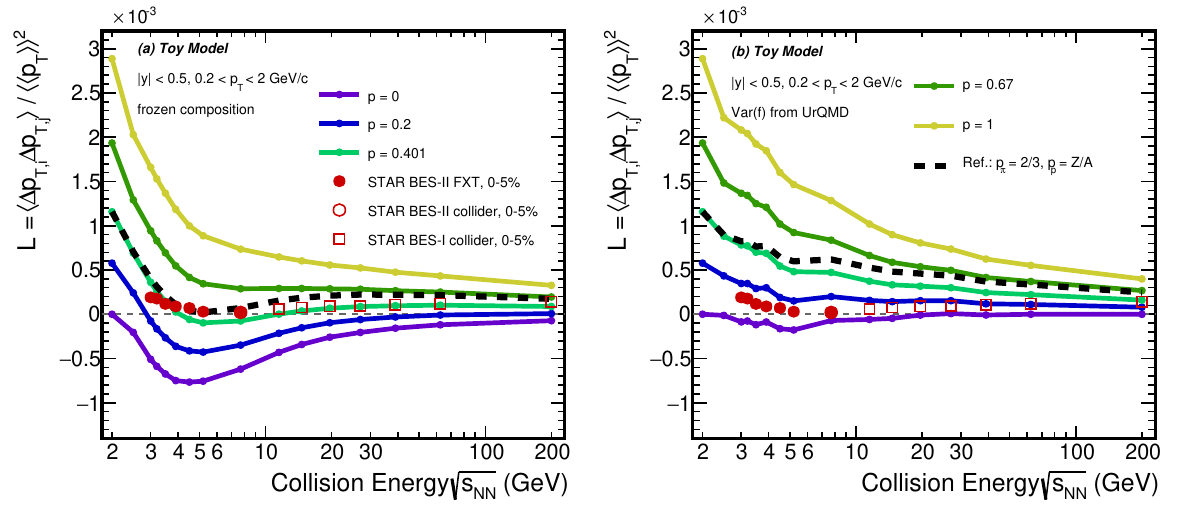}
\caption{The pairing probability $p$ swept from $0$ to $1$, with the
composition frozen (a) and fluctuating as measured in UrQMD (b), against the
measurement~\cite{STAR:2026vjv}.}
\label{fig:cons}
\end{figure*}

The pairing probability is not derived by Reichert \textit{et al.}; it is
the one adjustable parameter of that calculation. Figure~\ref{fig:cons} sweeps
it from $0$ to $1$ in both composition scenarios. The term acts only on
$\mathrm{Var}(\langle p_{\rm T}\rangle_k)$ and leaves
$\langle s^{2}_{p_T,k}/(N_k-1)\rangle$ untouched, as Fig.~2 of Reichert \textit{et al.} also shows, so it is an additive and monotonic
contribution: it can raise a curve but cannot introduce structure into one.
With the composition frozen it lifts a curve that already possesses a
minimum; with the composition fluctuating there is no minimum for it to lift,
and none appears at any value of the probability.

Once the composition is allowed to fluctuate as measured, the pairing
probability that would be needed to reproduce the data is small at every
energy, between $0.03$ and $0.13$, well below the $Z/A=0.401$ adopted by Reichert \textit{et al.}: with a fluctuating composition the mixing
requires almost no momentum conservation to reach the measured values, and
what the parameter then sets is the normalisation rather than the shape.

\subsection{Choice of acceptance}

\begin{figure*}[htbp]
\centering
\includegraphics[width=0.92\textwidth]{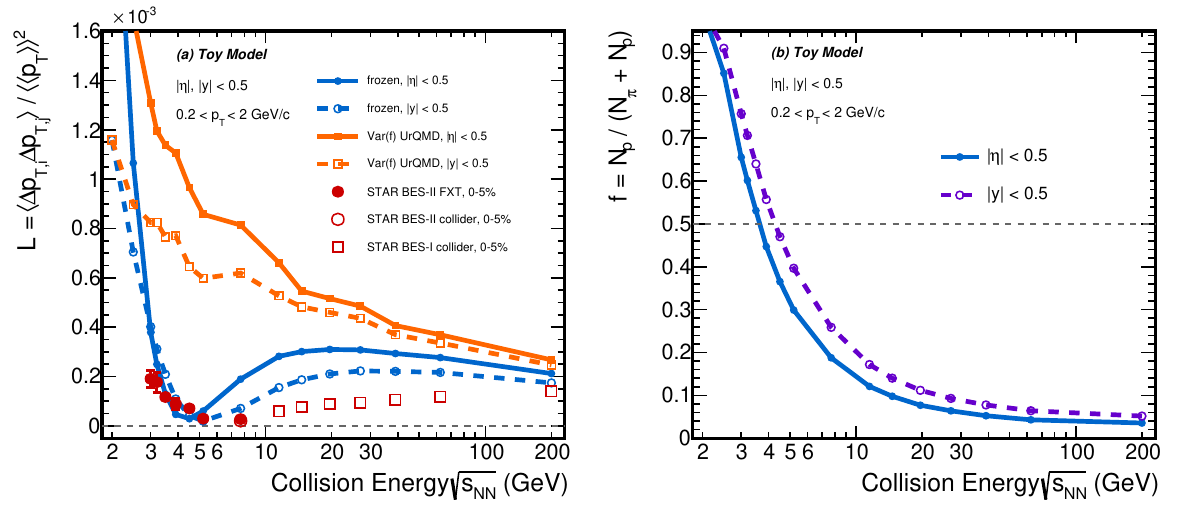}
\caption{(a) The two-component contribution in both acceptances, for frozen
and for measured composition fluctuation. (b) The baryon fraction the
two-species model sees, $f=N_p/(N_\pi+N_p)$, in the two windows; the arrows
mark equal abundance.}
\label{fig:acc}
\end{figure*}

The measurement is made in $|\eta|<0.5$; Reichert \textit{et al.} take their yields in $|y|<0.5$. For heavy particles the two are not
interchangeable, since at low transverse momentum a proton has $|\eta|>|y|$,
so the accepted proton-to-pion ratio differs between them. Because the
composition contribution is controlled by $f_B(1-f_B)$, this moves the
minimum: in the model's own two-species counting the abundances are equal at
$3.6$~GeV in $|\eta|<0.5$ and at $4.3$~GeV in $|y|<0.5$, and the minimum of
the frozen-composition curve moves correspondingly from $4.2$ to $5.6$~GeV.
Figure~\ref{fig:acc} shows both. The shift is a little over one~GeV and does
not affect the comparison of Fig.~\ref{fig:mixing}, in which both curves are
evaluated in the same acceptance; we record it because the position of the
minimum has been quoted in the literature without the acceptance being
stated alongside it.

\subsection{Tabulated inputs}

Tables~\ref{tab:inputs_eta} and~\ref{tab:inputs_y} list everything the
two-component model takes from transport, so that the calculation can be
repeated without rerunning UrQMD.

\begin{table*}[htbp]
\caption{Inputs to the two-component model, taken from UrQMD~3.4 cascade Au+Au events in the $0$--$5\%$ centrality class and the acceptance of the measurement, $|\eta|<0.5$ and $0.2<p_T<2$~GeV/$c$. $f_B=\langle N_p\rangle/\langle N\rangle$ is the baryon fraction of the accepted charged particles, $\Delta=\mu_B-\mu_M$ the difference of the species mean transverse momenta, and the last column the composition-fluctuation spectrum expressed as a ratio to the binomial value $f_B(1-f_B)/N$. Uncertainties are bootstrap over events; $10^4$ minimum-bias events per energy, of which approximately $500$ fall in the $0$--$5\%$ class. At $\sqrt{s_{NN}}=2.0$~GeV the accepted sample contains essentially no pions, so $f_B(1-f_B)$ vanishes and the composition mechanism is inoperative; the ratio quoted in the last column is then evaluated on a degenerate quantity and carries no information.}
\label{tab:inputs_eta}
\begin{ruledtabular}
\begin{tabular}{r r r r r c c c}
$\sqrt{s_{NN}}$ & $\langle N_\pi\rangle$ & $\langle N_K\rangle$ & $\langle N_p\rangle$ & $\langle N\rangle$ & $f_B$ & $\Delta$ & $\mathrm{Var}(f)/\mathrm{bin.}$ \\
(GeV) & & & & & & (GeV/$c$) & \\
\colrule
2.0 & 0.0 & 0.0 & 37.1 & 37.1 & 1.000 & 0.118 & $0.936 \pm 0.109$ \\
2.5 & 8.8 & 0.1 & 50.4 & 59.4 & 0.850 & 0.304 & $0.988 \pm 0.063$ \\
3.0 & 25.2 & 1.5 & 47.9 & 74.7 & 0.642 & 0.355 & $0.864 \pm 0.058$ \\
3.2 & 31.5 & 2.4 & 47.5 & 81.5 & 0.584 & 0.355 & $0.840 \pm 0.059$ \\
3.5 & 40.3 & 3.7 & 45.6 & 89.6 & 0.510 & 0.370 & $0.859 \pm 0.053$ \\
3.9 & 53.2 & 5.1 & 43.0 & 101.3 & 0.425 & 0.377 & $0.915 \pm 0.052$ \\
4.5 & 70.4 & 7.1 & 40.5 & 118.0 & 0.343 & 0.385 & $0.890 \pm 0.058$ \\
5.2 & 86.1 & 9.0 & 36.7 & 131.8 & 0.278 & 0.392 & $0.844 \pm 0.050$ \\
7.7 & 126.1 & 13.7 & 29.1 & 169.0 & 0.172 & 0.400 & $0.969 \pm 0.058$ \\
11.5 & 172.6 & 17.9 & 23.7 & 214.1 & 0.110 & 0.409 & $0.958 \pm 0.062$ \\
14.6 & 201.3 & 21.1 & 21.8 & 244.2 & 0.089 & 0.414 & $0.815 \pm 0.052$ \\
19.6 & 236.8 & 24.6 & 19.7 & 281.1 & 0.070 & 0.418 & $0.944 \pm 0.060$ \\
27.0 & 267.4 & 27.4 & 18.2 & 313.0 & 0.058 & 0.411 & $1.051 \pm 0.074$ \\
39.0 & 309.8 & 32.2 & 17.1 & 359.1 & 0.048 & 0.403 & $0.906 \pm 0.055$ \\
62.4 & 356.1 & 37.2 & 15.9 & 409.2 & 0.039 & 0.422 & $1.004 \pm 0.066$ \\
200.0 & 494.2 & 55.5 & 18.1 & 567.9 & 0.032 & 0.435 & $0.980 \pm 0.057$ \\
\end{tabular}
\end{ruledtabular}
\end{table*}

\begin{table*}[htbp]
\caption{As Table~\ref{tab:inputs_eta}, in the mid-rapidity acceptance $|y|<0.5$ used by Reichert \textit{et al.}~\cite{Reichert:2026mbm}.}
\label{tab:inputs_y}
\begin{ruledtabular}
\begin{tabular}{r r r r r c c c}
$\sqrt{s_{NN}}$ & $\langle N_\pi\rangle$ & $\langle N_K\rangle$ & $\langle N_p\rangle$ & $\langle N\rangle$ & $f_B$ & $\Delta$ & $\mathrm{Var}(f)/\mathrm{bin.}$ \\
(GeV) & & & & & & (GeV/$c$) & \\
\colrule
2.0 & 0.0 & 0.0 & 80.1 & 80.1 & 1.000 & 0.090 & $0.935 \pm 0.045$ \\
2.5 & 9.5 & 0.2 & 95.7 & 105.4 & 0.908 & 0.225 & $0.935 \pm 0.051$ \\
3.0 & 27.1 & 2.1 & 84.4 & 113.6 & 0.744 & 0.264 & $0.830 \pm 0.059$ \\
3.2 & 33.8 & 3.3 & 81.3 & 118.4 & 0.688 & 0.267 & $0.870 \pm 0.057$ \\
3.5 & 43.4 & 4.9 & 77.1 & 125.4 & 0.616 & 0.280 & $0.825 \pm 0.051$ \\
3.9 & 56.9 & 7.0 & 71.5 & 135.4 & 0.529 & 0.286 & $0.883 \pm 0.054$ \\
4.5 & 75.0 & 9.8 & 66.5 & 151.3 & 0.440 & 0.295 & $0.787 \pm 0.049$ \\
5.2 & 91.8 & 12.5 & 60.3 & 164.6 & 0.367 & 0.303 & $0.765 \pm 0.048$ \\
7.7 & 134.5 & 18.8 & 46.9 & 200.1 & 0.234 & 0.312 & $0.885 \pm 0.052$ \\
11.5 & 183.7 & 24.8 & 38.1 & 246.7 & 0.155 & 0.318 & $0.863 \pm 0.063$ \\
14.6 & 214.4 & 29.4 & 34.9 & 278.7 & 0.125 & 0.328 & $0.864 \pm 0.061$ \\
19.6 & 252.3 & 34.1 & 31.7 & 318.2 & 0.100 & 0.327 & $0.962 \pm 0.060$ \\
27.0 & 284.7 & 38.1 & 29.1 & 351.9 & 0.083 & 0.325 & $1.036 \pm 0.059$ \\
39.0 & 329.9 & 44.8 & 27.7 & 402.4 & 0.069 & 0.316 & $0.945 \pm 0.055$ \\
62.4 & 379.0 & 51.9 & 25.9 & 456.7 & 0.057 & 0.324 & $0.999 \pm 0.063$ \\
200.0 & 524.9 & 76.3 & 28.5 & 629.7 & 0.045 & 0.345 & $0.994 \pm 0.060$ \\
\end{tabular}
\end{ruledtabular}
\end{table*}

\begin{table*}[htbp]
\caption{Reproduction of the two-component calculation of Reichert \textit{et al.}~\cite{Reichert:2026mbm} in their own acceptance ($|y|<0.5$, $0.2<p_T<2$~GeV/$c$), with frozen composition and the effective temperatures and pairing probabilities quoted there, for the three momentum-conservation scenarios of its Fig.~1. Values are $\mathcal{L}$, normalised as in Eq.~(\ref{eq:linear}); uncertainties are the standard error over independent evaluations.}
\label{tab:repro}
\begin{ruledtabular}
\begin{tabular}{r c c c}
$\sqrt{s_{NN}}$ & no conservation & partial pairing & full pairing \\
(GeV) & ($10^{-4}$) & ($10^{-4}$) & ($10^{-4}$) \\
\colrule
2.0 & $+0.00 \pm 0.04$ & $+11.70 \pm 0.03$ & $+28.97 \pm 0.07$ \\
2.5 & $-2.11 \pm 0.02$ & $+6.92 \pm 0.02$ & $+20.03 \pm 0.04$ \\
3.0 & $-5.09 \pm 0.03$ & $+3.96 \pm 0.02$ & $+16.59 \pm 0.04$ \\
3.2 & $-5.97 \pm 0.01$ & $+2.97 \pm 0.02$ & $+15.05 \pm 0.04$ \\
3.5 & $-6.73 \pm 0.03$ & $+1.97 \pm 0.02$ & $+13.39 \pm 0.02$ \\
3.9 & $-7.54 \pm 0.02$ & $+0.91 \pm 0.01$ & $+11.55 \pm 0.02$ \\
4.5 & $-7.71 \pm 0.02$ & $+0.34 \pm 0.02$ & $+9.94 \pm 0.02$ \\
5.2 & $-7.56 \pm 0.01$ & $+0.20 \pm 0.02$ & $+8.86 \pm 0.02$ \\
7.7 & $-6.25 \pm 0.01$ & $+0.62 \pm 0.02$ & $+7.15 \pm 0.02$ \\
11.5 & $-4.30 \pm 0.01$ & $+1.54 \pm 0.01$ & $+6.47 \pm 0.02$ \\
14.6 & $-3.45 \pm 0.01$ & $+1.82 \pm 0.01$ & $+5.93 \pm 0.01$ \\
19.6 & $-2.61 \pm 0.01$ & $+2.09 \pm 0.01$ & $+5.53 \pm 0.00$ \\
27.0 & $-2.06 \pm 0.01$ & $+2.20 \pm 0.01$ & $+5.13 \pm 0.01$ \\
39.0 & $-1.58 \pm 0.01$ & $+2.19 \pm 0.01$ & $+4.75 \pm 0.02$ \\
62.4 & $-1.20 \pm 0.01$ & $+2.16 \pm 0.01$ & $+4.31 \pm 0.01$ \\
200.0 & $-0.74 \pm 0.00$ & $+1.72 \pm 0.01$ & $+3.23 \pm 0.00$ \\
\end{tabular}
\end{ruledtabular}
\end{table*}

\section{Relations between mean-\texorpdfstring{$p_{\rm T}$}{pT} fluctuation measures}
\label{app:measures}

Measurements of event-by-event mean transverse-momentum fluctuations have been reported in several different variables. This appendix collects the relations between them, so that published values can be placed on a common scale.

Let $N$ be the accepted charged-particle multiplicity in an event, $\langle N\rangle$ its mean over the centrality class, and $\sigma^{2}_{p_T}$ the variance of the inclusive transverse-momentum distribution within the same acceptance. Writing $[p_{\rm T}]_k$ for the event-wise mean, the dynamical variance introduced in Ref.~\cite{Voloshin:1999yf} is
\begin{equation}
\sigma^{2}_{\mathrm{dyn}}
=\mathrm{Var}\!\left([p_{\rm T}]_k\right)-\frac{\sigma^{2}_{p_T}}{\langle N\rangle}
=\langle \Delta p_{T,i}\,\Delta p_{T,j}\rangle ,
\label{eq:app-dyn}
\end{equation}
identical to Eq.~(\ref{eq:decomp}) up to the distinction between $N$ and $N-1$, which is immaterial at the multiplicities considered here. It is convenient to define the dimensionless ratio
\begin{equation}
R=\frac{\langle N\rangle\,\langle \Delta p_{T,i}\,\Delta p_{T,j}\rangle}{\sigma^{2}_{p_T}} .
\label{eq:app-R}
\end{equation}

The measures then relate as follows.

\emph{Scaled correlator and linear variable} (STAR~\cite{STAR:2019clv,STAR:2026vjv}, and this work):
\begin{equation}
C_{p_T}=\frac{\sqrt{\langle \Delta p_{T,i}\Delta p_{T,j}\rangle}}{\langle\langle p_{\rm T}\rangle\rangle},
\qquad
\mathcal{L}=C_{p_T}^{2} .
\end{equation}

\emph{ATLAS cumulant} $k_2$~\cite{PhysRevLett.133.252301}: the correlator is constructed with the same inclusive-mean subtraction and pair normalisation, and is divided by $\langle [p_{\rm T}]\rangle^{2}$, so that
\begin{equation}
k_2=\mathcal{L}
\end{equation}
identically, with no approximation and no auxiliary quantities required.

\emph{Relative dynamical fluctuation} $F_{p_T}$ (PHENIX~\cite{PHENIX:2003ccl}):
\begin{equation}
F_{p_T}=\sqrt{1+R}-1 \;\simeq\; \frac{R}{2}\quad (R\ll 1).
\end{equation}

\emph{Strongly intensive measure} $\Phi_{p_T}$ (introduced in Ref.~\cite{Gazdzicki:1992ri}; used by NA49~\cite{NA49:1999inh} and CERES~\cite{Appelshauser:2004xj}):
\begin{equation}
\begin{split}
\Phi_{p_T}&=\sigma_{p_T}\left(\sqrt{1+R}-1\right)=\sigma_{p_T}\,F_{p_T}\\
&\simeq\frac{\langle N\rangle\,\langle \Delta p_{T,i}\Delta p_{T,j}\rangle}{2\,\sigma_{p_T}} .
\end{split}
\end{equation}

Two practical consequences follow. First, $C_{p_T}$, $\mathcal{L}$ and $k_2$ are interconvertible using nothing beyond the measurement itself, whereas recovering $\langle \Delta p_{T,i}\Delta p_{T,j}\rangle$ from a published $\Phi_{p_T}$ or $F_{p_T}$ additionally requires $\langle N\rangle$ and $\sigma_{p_T}$ evaluated in the same acceptance. Second, and more importantly, all of these quantities are acceptance dependent: the pseudorapidity window sets $\langle N\rangle$, and the transverse-momentum cuts set both $\sigma_{p_T}$ and $\langle\langle p_{\rm T}\rangle\rangle$. Values measured in different acceptances are therefore not directly comparable, and any compilation of them should be read with the acceptances of Table~\ref{tab:acceptances} in mind.

\begin{table*}[htbp]
\caption{Kinematic acceptance and reported variable for the measurements discussed in this work. The ISR, CERES, ALICE and STAR entries are those discussed in the Conclusion; the NA49, PHENIX and ATLAS measurements are quoted in variables related to $C_{p_T}$ as set out in Appendix~\ref{app:measures}.}
\label{tab:acceptances}
\begin{ruledtabular}
\begin{tabular}{llll}
Experiment & System & Acceptance & Variable \\
\colrule
ISR    & $p+p$   & $|y|<0.8, 0.15<p_{T}<2.0~GeV/c$                                        & $\sigma(\langle p_{\rm T}\rangle)$ \\
CERES  & Pb+Au   & $2.2 <\eta<2.7$, $0.1 <p_{T}<1.5~GeV/c$
                                       & $\Sigma_{p_T}$ \\
NA49   & Pb+Pb   & $4<y<5.5$, $0.005<p_{T}<1.5$~GeV$/c$       & $\Phi_{p_T}$ \\
PHENIX & Au+Au   & $|\eta|<0.35$, $0.2<p_{T}<2.0$~GeV$/c$     & $F_{p_T}$ \\
STAR   & Au+Au   & $|\eta|<0.5$, $0.2<p_{T}<2.0$~GeV$/c$      & $C_{p_T}$ \\
ALICE  & Pb+Pb, Xe+Xe, $p+p$ & $|\eta|<0.8$, $0.15<p_{T}<2.0$~GeV$/c$ & $C_{p_T}$ \\
ATLAS  & Pb+Pb, Xe+Xe        & $|\eta|<2.5$, $p_{T}>0.5$~GeV$/c$      & $k_2$ \\
\end{tabular}
\end{ruledtabular}
\end{table*}

The ATLAS acceptance differs substantially from the others --- a five times wider pseudorapidity window and a transverse-momentum threshold that removes the soft region entirely --- so those measurements are not directly comparable with the others.

\bibliographystyle{apsrev4-2} 
\bibliography{References}

\end{document}